\documentclass[lettersize,journal]{IEEEtran}
\usepackage{amsmath,amsfonts}
\usepackage{array}
\usepackage[caption=false,font=normalsize,labelfont=sf,textfont=sf]{subfig}
\usepackage{textcomp}
\usepackage{stfloats}
\usepackage{url}
\usepackage{verbatim}
\usepackage[switch]{lineno}
\usepackage{graphicx}

\usepackage{cite}
\usepackage{amsmath,amssymb,amsfonts}

\usepackage{algorithm}
\usepackage{algpseudocode}
\usepackage{xcolor}
\algrenewcommand\alglinenumber[1]{\scriptsize #1:}

\newcommand{\algcmt}[1]{\hspace{0.5em}\textcolor{blue}{// #1}}

\usepackage{graphicx}
\usepackage{textcomp}
\usepackage{xcolor}
\usepackage{booktabs}
\usepackage{multirow}
\usepackage{url}
\usepackage[hidelinks]{hyperref}

\usepackage{amssymb}
\usepackage{xcolor} 
\usepackage[switch]{lineno} 

\graphicspath{{paper_draw/}}

\newcommand{\method}{UnionSparse}
\newcommand{\vdir}{\textsc{PMR}}
\newcommand{\ci}{\textsc{CI}}

\newcommand{\lspd}{\textsc{LSPD}}

\def\BibTeX{{\rm B\kern-.05em{\sc i\kern-.025em b}\kern-.08em
    T\kern-.1667em\lower.7ex\hbox{E}\kern-.125emX}}
\usepackage{balance}

\begin{document}
\IEEEpubid{\parbox{\textwidth}{\centering\tiny \textcopyright{} 2026 IEEE. Personal use of this material is permitted. Permission from IEEE must be obtained for all other uses, in any current or future media, including reprinting/republishing this material for advertising or promotional purposes, creating new collective works, for resale or redistribution to servers or lists, or reuse of any copyrighted component of this work in other works.}}
\title{UnionSparse: An Index-Efficient Sparsity Framework for Low-Bit Sparse  LLM Inference on Edge}
\author{Tianhao~Jiang\textsuperscript{*}\thanks{\textsuperscript{*} Tianhao Jiang and Hang Gu contributed equally to this work.}, Hang~Gu\textsuperscript{*}, Teng~Wang, Qianyu~Cheng, ZhenDong~Zheng, Cheng~Tang, Qiyue~Su, Wenqi~Lou, Lei~Gong, Chao~Wang, Xi~Li, and Xuehai~Zhou%
\thanks{Tianhao Jiang, Hang Gu, Teng Wang, Qianyu Cheng, ZhenDong Zheng, Cheng Tang, Qiyue Su, Wenqi Lou, Lei Gong, Chao Wang, Xi Li, and Xuehai Zhou are with the University of Science and Technology of China, Hefei, China (e-mail: thjiang@mail.ustc.edu.cn; hanggu@mail.ustc.edu.cn; wangt635@ustc.edu.cn; qycheng@mail.ustc.edu.cn; zzd1411@mail.ustc.edu.cn; sisyphustc@mail.ustc.edu.cn; qiyuesu@mail.ustc.edu.cn; louwenqi@ustc.edu.cn; leigong0203@ustc.edu.cn; cswang@ustc.edu.cn; llxx@ustc.edu.cn; xhzhou@ustc.edu.cn).}}

\markboth{IEEE Transactions on Computer-Aided Design of Integrated Circuits and Systems}%
{Jiang \MakeLowercase{\textit{et al.}}: UnionSparse}
\maketitle

\begin{abstract}
Edge LLM inference combines sparsity and low-bit quantization to meet device memory, latency, and power limits. Yet quantization shrinks weight payloads without proportionally reducing sparse metadata, so index traffic and nonzero extraction become critical SpMM bottlenecks. We introduce the Payload-to-Metadata Ratio (\vdir) and show that improving \vdir{} raises effective compute intensity in decoding.

We present UnionSparse, an index-efficient framework that combines Index-Efficient Bitmap Encoding (IE-BME) with a SpMM kernel using Low-Bit Shared-Memory Parallel Decoding (LSPD). IE-BME amortizes metadata and aligns sparse traversal with fragment assembly, while LSPD improves small-batch execution. Under W4A4 quantization and 30\%--70\% sparsity, UnionSparse outperforms FlashLLM and SpInfer by 2.30$\times$ and 1.43$\times$, and CUTLASS and cuBLAS Tensor Core by 1.56$\times$ and 3.46$\times$, respectively. These results establish payload-extraction efficiency as a first-order concern for low-bit sparse inference on edge GPUs. Source code is available at:
\url{https://github.com/Victor-Alen/UnionSparse}.
\end{abstract}


\section{Introduction}

\begin{figure}[t]
    \centering
\includegraphics[width=0.76\linewidth, trim=12pt 15pt 12pt 10pt, clip]{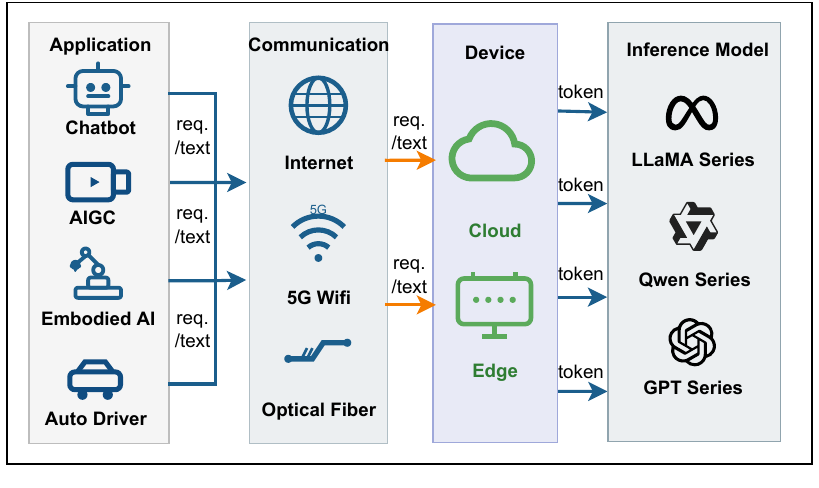}
    \caption{Representative LLM deployment scenarios and service workflow.}
    \label{fig:bk}
\end{figure}

\IEEEPARstart{L}{arge} language models (LLMs) are moving into embodied agents, industrial systems, and vehicles~\cite{transformer,llama2,gpt4}, as illustrated in Fig.~\ref{fig:bk}. Their tight memory, power, and thermal budgets~\cite{xu2025edgellm,huang2025tenet,cai2026efficient} make small-batch autoregressive decoding, with repeated projection and feed-forward layers, a central edge-inference challenge.

Sparsification~\cite{gopalan2026hqp,husom2025sustainable,wang2025unified,cai2026efficient} removes redundant weights, while quantization~\cite{li2025veclut,fu2025h2eal} compresses weights and activations into formats such as INT8, FP8, and INT4~\cite{gptq,awq,qserve,flatquant,llmint8,smoothquant,sparsegpt,tcad24_naf}. Their combination reduces both computation and memory traffic for edge deployment.

Efficient low-bit sparse inference nevertheless remains difficult. SpMM dominates QKV, attention-output, and feed-forward projections; during small-$N$ decoding, their tall-and-skinny shapes have low arithmetic intensity and are constrained by data movement and sparse decoding rather than arithmetic throughput.

Quantization shrinks weight payloads but leaves conventional sparse metadata largely unchanged. At lower precision, metadata occupies more traffic and, with little decode-time reuse, its movement and decoding may dominate despite reduced arithmetic. Efficiency therefore depends not only on the number of nonzeros, but also on how they are represented, indexed, traversed, and assembled.

Existing GPU SpMM designs improve scheduling, Tensor Core utilization, and runtime integration, but edge GPUs expose less bandwidth and parallelism than server GPUs. Metadata overhead and irregular traversal thus motivate treating index efficiency as a first-order design objective.

We formalize this bottleneck with the \textit{Payload-to-Metadata Ratio} (\vdir) and relate it to sparse-kernel \textit{compute intensity} (\ci), explaining why pruning may leave low-bit kernels bandwidth-bound and why index efficiency matters increasingly at lower bit widths.

We present \method{}\footnote{https://github.com/Victor-Alen/UnionSparse}, which co-designs sparse representation and execution. Index-Efficient Bitmap Encoding (IE-BME) improves \vdir{} and regularizes traversal; an optimized SpMM pipeline combines Low-Bit Shared-Memory Parallel Decoding (LSPD), asynchronous movement, and hardware-aligned fragment assembly for Tensor Core execution.

Our contributions are summarized as follows:
\begin{itemize}

\item We provide a theoretical analysis showing that indexing overhead and irregular memory access are two principal factors limiting the performance of low-bit SpMM under small-batch inference on edge.

\item We propose \method{} and its IE-BME format, which improves \ci{} while aligning sparse traversal, fragment assembly, and Tensor Core execution.

\item We design and implement Low-Bit Shared-Memory Parallel Decoding (\lspd), which unifies the proposed representation, asynchronous data movement, and shape-aware kernel selection into a cohesive execution framework.

\item We demonstrate that \method{} substantially outperforms state-of-the-art dense and sparse baselines at both kernel and framework levels through evaluation under W4A4 quantization on edge GPUs.

\end{itemize}

\section{Background and Related Work}

\subsection{LLM Inference on Edge}

Foundation models are moving from datacenters to robots, autonomous systems, personal assistants, and private servers~\cite{driess2023palm,gao2026foundation,aas2023intelligent,knoop2026private}. Local execution can reduce latency and cloud cost while improving privacy and connectivity tolerance.

Autoregressive inference has \textit{prefill} and \textit{decode} stages~\cite{transformer}. Prefill processes prompt tokens in parallel to build hidden states and the key-value cache; decode generates one or a few tokens while reusing that cache~\cite{kwon2023efficient} and directly determines token latency.

Edge memory, bandwidth, power, and thermal limits often make dense FP16/BF16 execution impractical~\cite{gptq,smoothquant}. Pruning and quantization reduce model size, traffic, and arithmetic cost, enabling larger models on constrained devices~\cite{sparsegpt,sparsity_survey,tan2024wrp,awq,zhang2024structured,wang2025unisparta,tang2026closertome}.

\subsection{SpMM Workloads}

\begin{figure}[t]
    \centering
    \includegraphics[width=0.83\linewidth]{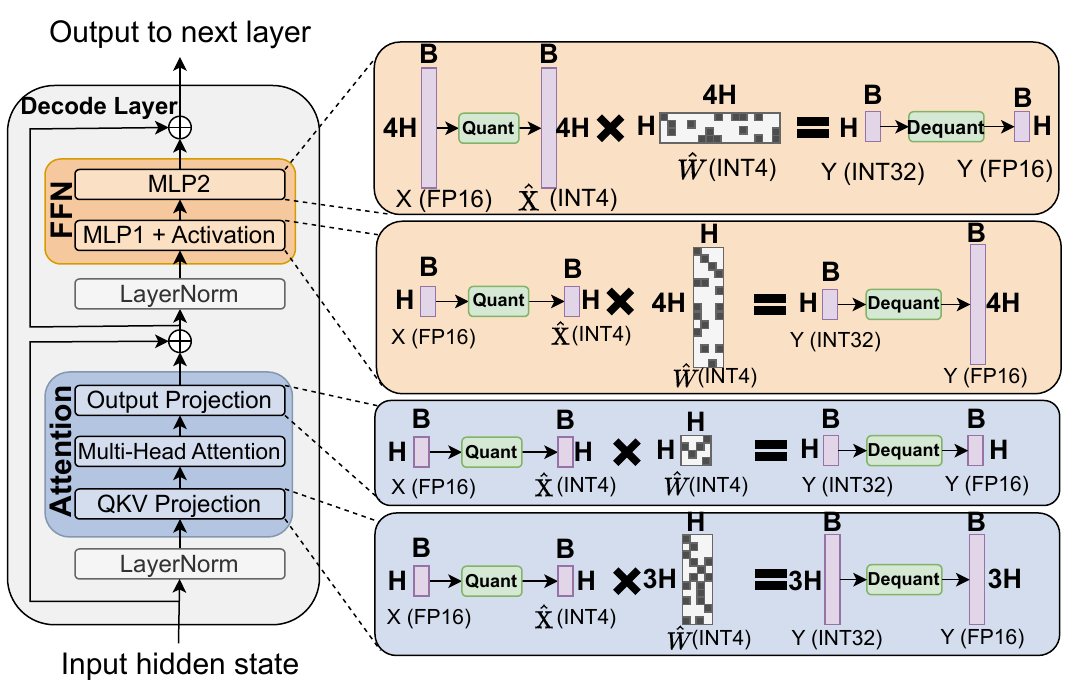}
    \caption{The four major matrix multiplications in the Transformer architecture. Here, $H$ denotes the hidden size and $B$ denotes the batch size. In the weight matrix $\hat{W}$, the darker elements indicate the weights retained after pruning.}
    \label{fig:fourmm}
\end{figure}




Figure~\ref{fig:fourmm} identifies four dominant linear operators: QKV and attention-output projections and the two FFN projections. They appear in both inference stages, but prefill exposes token-level parallelism whereas decode processes few new tokens and is therefore more latency- and bandwidth-sensitive.

\begin{figure}[t]
  \centering
  \includegraphics[width=0.83\linewidth]{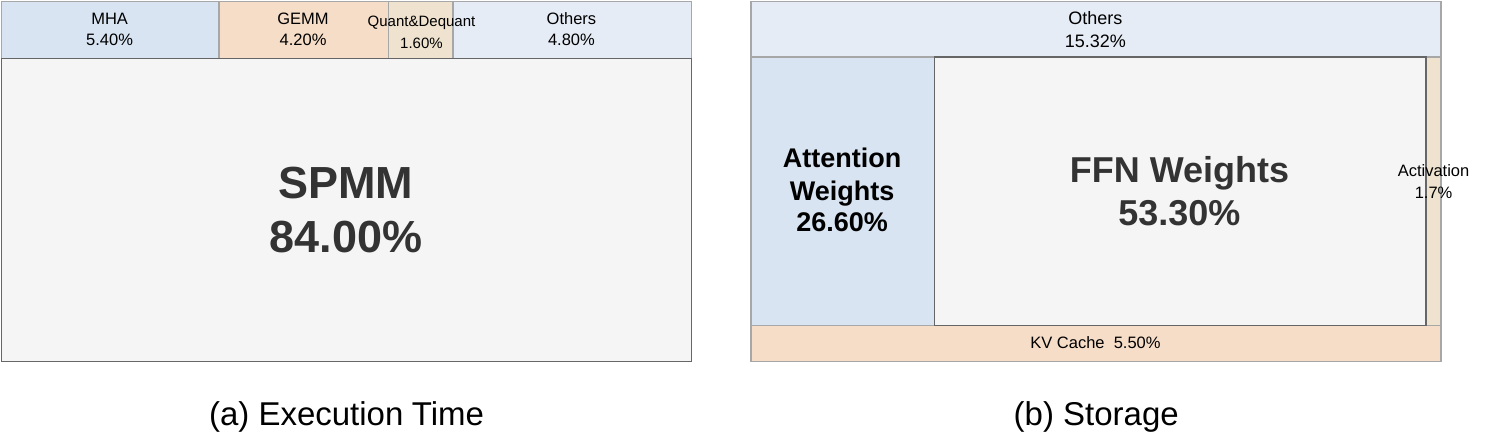}
  \caption{Breakdown of OPT-13B inference on Jetson AGX Orin 64GB.}
  \label{fig:opt13b-breakdown}
\end{figure}

Profiling OPT-13B on Jetson AGX Orin 64GB in Fig.~\ref{fig:opt13b-breakdown} confirms that SpMM dominates runtime, while FFN and attention weights dominate storage. These matrix-heavy operators are therefore the primary compression and execution targets.

With pruning and quantization, these layers map to low-bit sparse-dense kernels. Their efficiency depends on retained computation and on how compressed data are indexed, moved, and assembled, especially in bandwidth-sensitive decode where metadata is difficult to amortize.

\subsection{Related Work}
Sparse GPU work comprises general-purpose kernels and LLM-oriented inference frameworks.

\textbf{General-Purpose SpMM.} General-purpose SpMM studies sparse-dense multiplication as a reusable sparse linear algebra primitive rather than a workload tied to a specific model family. Early sparse GPU kernels for deep learning, such as Sputnik \cite{gale2020sparse}, already demonstrated that CSR-based sparse matrix multiplication can substantially accelerate pruned DNN models on GPUs, making sparse SpMM a practical building block beyond traditional scientific computing. Representative kernel-level works, including DTC-SpMM \cite{fan2024dtc}, SMaT \cite{okanovic2024smat}, Acc-SpMM \cite{zhao2025acc}, and Voltrix-SpMM \cite{xia2025voltrix}, further improve performance on modern GPUs through Tensor-Core-aware formats, data reordering, asynchronous pipelines, and load-balancing strategies. Beyond individual kernels, SparTA \cite{zheng2022sparta} extends this line toward an end-to-end sparse deep learning framework, where sparsity attributes are propagated across the computation graph, and specialized sparse operators are generated accordingly. On the library side, NVIDIA cuSPARSE \cite{nvidia_cusparse} provides industrial-grade generic SpMM support via \texttt{cusparseSpMM()}, while cuSPARSELt \cite{nvidia_cusparselt} mainly targets structured sparse matrix multiplication on Sparse Tensor Cores. Overall, this line of work focuses on improving SpMM as a general sparse GPU primitive across scientific computing, graph analytics, and deep learning workloads.

\textbf{LLM-oriented SpMM.} 
More recent works have begun to specialize sparse GPU execution for large language model inference. Flash-LLM~\cite{flashllm} adopts a load-as-sparse, compute-as-dense design to reduce data movement while preserving dense Tensor Core execution. SpInfer~\cite{spinfer} further co-designs sparse representation and GPU execution through Tensor-Core-Aware Bitmap Encoding, shared-memory bitmap decoding, and asynchronous pipelining, improving both kernel-level SpMM efficiency and end-to-end sparse LLM inference. Compared with general-purpose sparse kernels, these methods are much better aligned with the execution characteristics of autoregressive Transformers, especially under unstructured pruning and skinny matrix shapes.

Edge GPUs have less bandwidth, cache capacity, and parallelism than server GPUs, making performance more sensitive to representation and metadata handling. Lower precision further increases relative index traffic, leaving index-efficient representation and execution as an open problem.

\section{Gap Analysis and Motivation}




\subsection{Bottlenecks in Edge LLM Inference}

We profile OPT-13B on Jetson AGX Orin 64GB with FasterTransformer, 50\% weight pruning, W4A4 quantization, batch size 2, and output length 512. In Fig.~\ref{fig:opt13b-breakdown}, FFN and attention weights occupy 53.3\% and 26.6\% of storage, while SpMM consumes 84\% of execution time, establishing sparse SpMM as the central target.

\subsection{Understanding Indexing Cost in Low-Bit Sparse LLM Inference}

Pruning and quantization help only if retained weights are extracted efficiently. Sparse payloads require offsets, indices, or bitmaps~\cite{spinfer,flashllm,sputnik}; quantization shrinks values but not this metadata proportionally.


To quantify the efficiency of payload extraction, we define the Payload-to-Metadata Ratio (\vdir) as
\begin{equation}
\small
\mathrm{PMR} = \frac{B_{w} \times M \times K \times (1-s)}{\mathrm{Stor}_{\text{meta}}}
\label{eq:vdir}
\end{equation}
where $B_w$ denotes bytes per weight element, $M$ and $K$ are the dimensions of the weight matrix, and $s$ denotes the sparsity. Accordingly, $MK(1-s)$ gives the number of nonzero elements ($NNZ$) remaining after pruning, and $B_w \cdot NNZ$ denotes the storage of useful payloads $\mathrm{Stor}_{\mathrm{payload}}$. $\mathrm{Stor}_{\mathrm{meta}}$ denotes the metadata storage required to locate these payloads. A higher PMR indicates that a more useful payload can be obtained per metadata access, whereas a lower PMR implies that metadata overhead becomes more dominant.

\begin{figure}[t]
  \centering
  \includegraphics[width=0.85\linewidth]{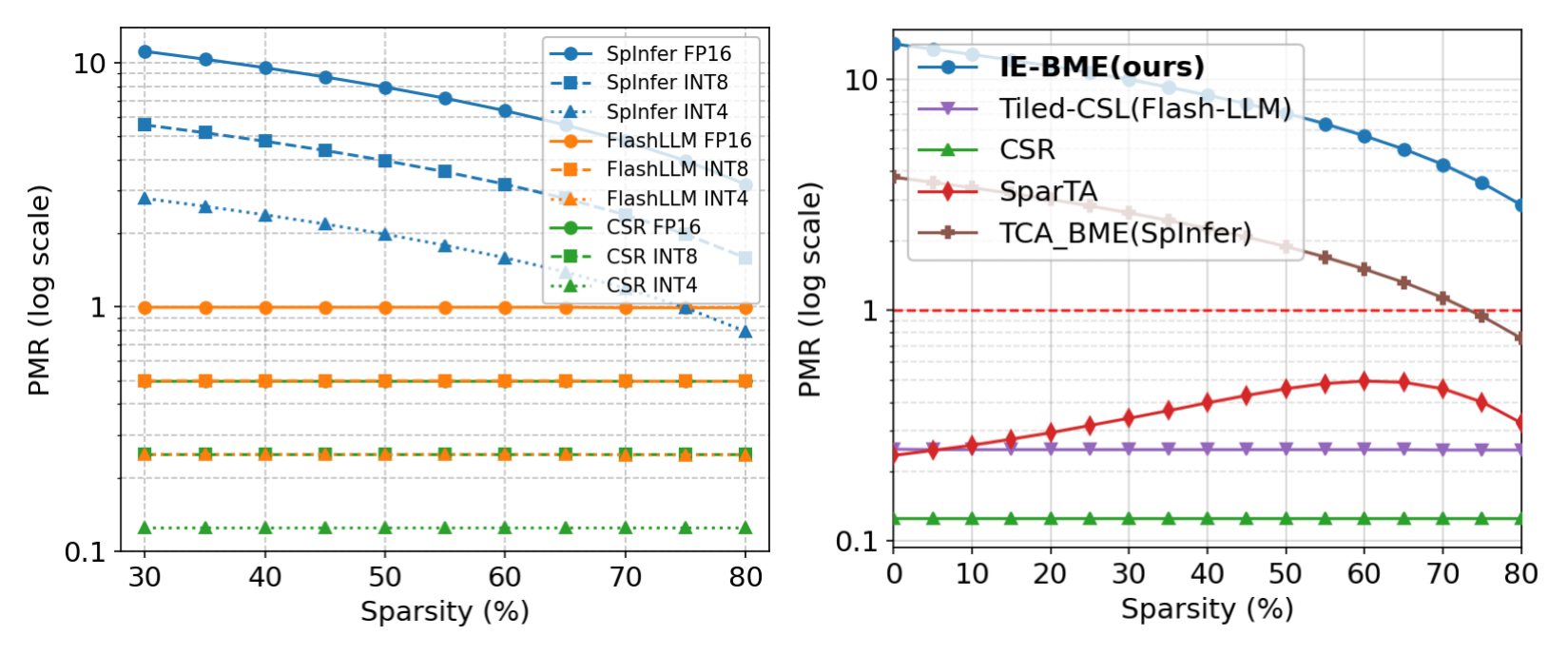}
\caption{PMR comparison across precision levels and sparse formats. The left panel compares FP16, INT8, and INT4 under three representative formats, while the right panel focuses on the INT4 regime and shows the PMR gap across different sparse representations.}
  \label{fig:pmr}
\end{figure}

Figure~\ref{fig:pmr} compares TCA-BME (SpInfer), Tiled-CSR (FlashLLM), and CSR. PMR decreases from FP16 to INT8 and INT4 for every format, showing that lower precision magnifies relative metadata cost; the persistent gap among formats shows that encoding design strongly affects payload efficiency. Thus, edge performance depends on both sparsity and maintaining high PMR at low precision.

\subsection{Roofline Analysis of PMR-Induced CI Shift}

For sparse low-bit SpMM, metadata overhead reduces the effective compute intensity seen by the hardware. Using \vdir, the effective CI can be written as
\begin{equation}
\small
{CI}_{\text{SpMM}}
=
\frac{2M N (1-s)}
{B_wM(1-s)\left(1+\frac{1}{\vdir}\right)+B_wN}
\end{equation}
where $B_w$ denotes bytes per weight element, the term \(\left(1+\frac{1}{\vdir}\right)\) captures the inflation of weight-side memory traffic caused by metadata overhead. As \vdir decreases, metadata accounts for a larger fraction of total traffic, thereby reducing the effective CI even when sparsity lowers arithmetic work.

\begin{figure}[t]
  \centering
  \includegraphics[width=0.85\linewidth]{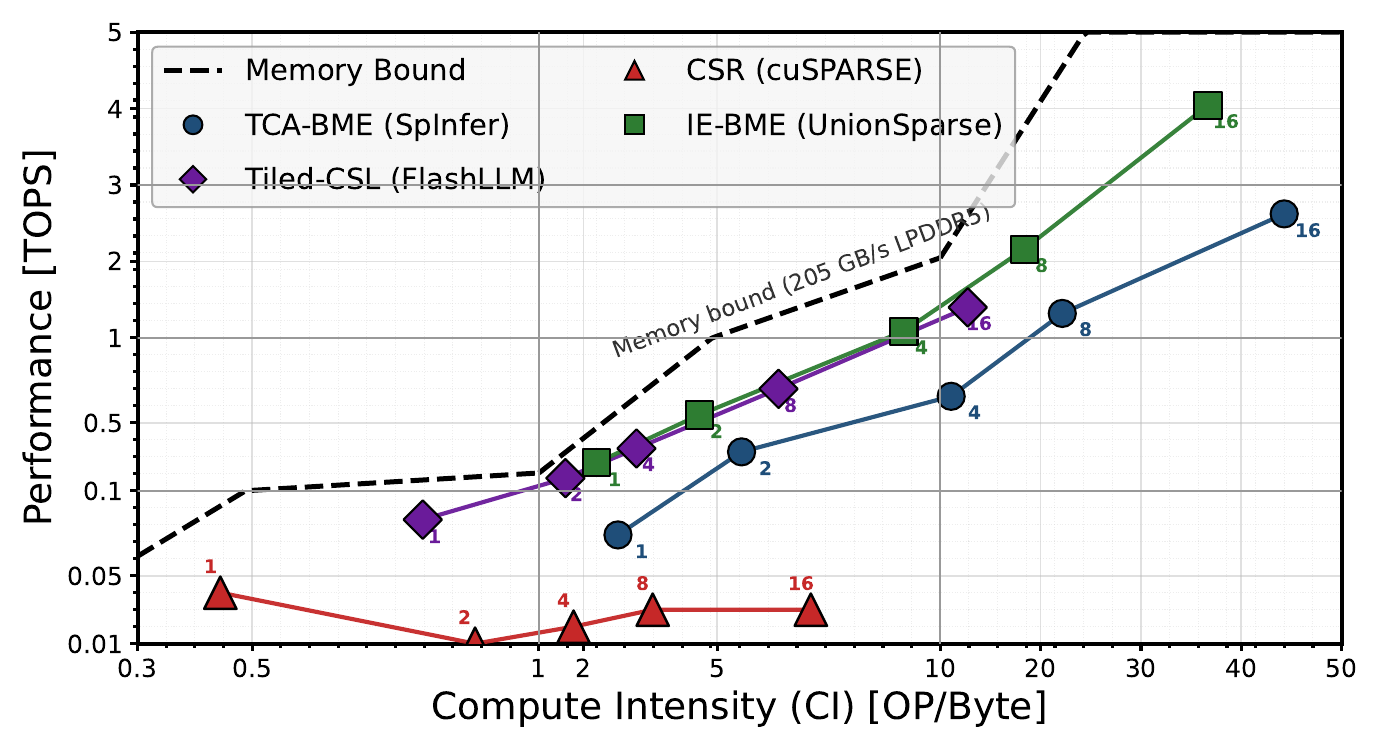}
\caption{Roofline view of different sparse formats under W4A4 quantization, with $m=28{,}672$, $k=8{,}192$, and sparsity of $40\%$. The x-axis shows the analytically computed effective CI, while the y-axis shows measured kernel performance. For the same marked $N$ value, higher-PMR formats shift rightward, indicating higher effective compute intensity and a greater ability to move away from the memory-bound region.}
  \label{fig:roofline}
\end{figure}

Figure~\ref{fig:roofline} plots the measured kernel performance against the analytically computed effective CI under W4A4 quantization. Here, the x-axis values are obtained from the above CI formulation, while the y-axis values are collected from actual kernel execution. The dashed line indicates the theoretical memory-bound roofline. 

Sparse formats with higher PMR consistently achieve higher effective CI and better measured performance. More importantly, for the same marked $N$ value, the operating points of high-PMR formats shift to the right, indicating that less metadata overhead is incurred for the same workload and more useful computation is delivered per byte of memory traffic. This rightward shift makes the kernel more likely to move away from the bandwidth-bound region and better exploit the available system performance. In contrast, low-PMR formats remain much closer to the bandwidth ceiling, so the theoretical arithmetic reduction from sparsity is harder to translate into a practical speedup.

Therefore, the key challenge of low-bit sparse LLM inference on edge is not merely to increase sparsity, but to improve index efficiency so that sparse kernels can sustain higher effective compute intensity and more readily escape the memory-bound regime.

\section{UnionSparse Design Overview}

\begin{figure*}[t]
    \centering
    \includegraphics[width=0.87\textwidth]{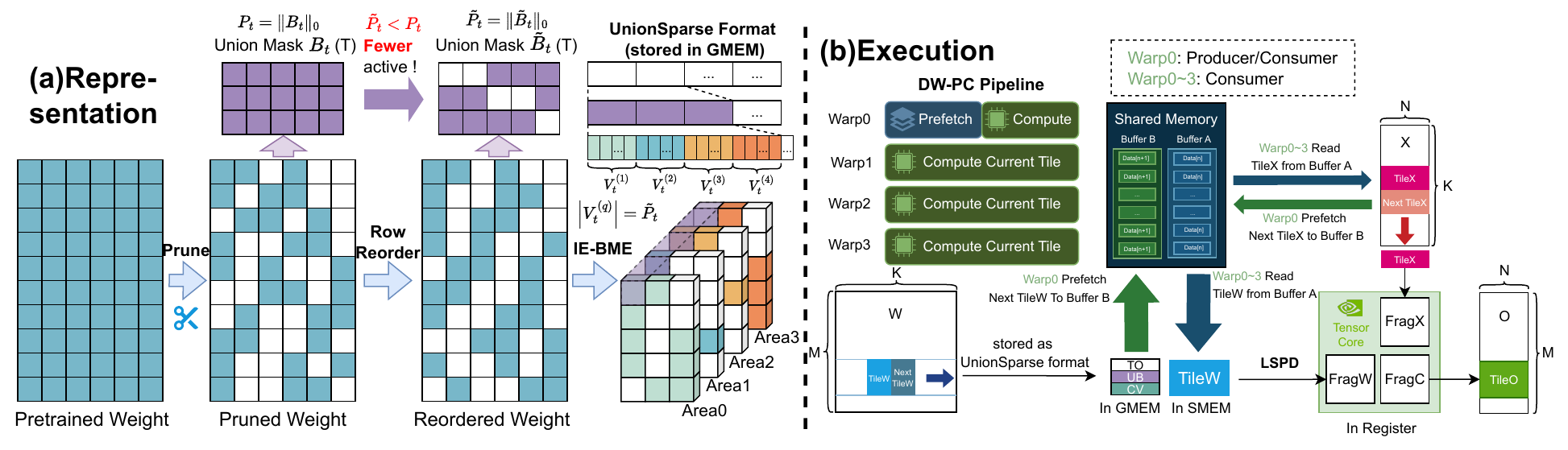}
    \caption{Overall framework of UnionSparse with the representation and execution optimization. The representation part reduces logical union-mask activity through row reordering and encodes the reordered weights using IE-BME, while the execution part combines the Decoupled Warp-Specialized Producer-Consumer (DW-PC) pipeline and Low-Bit Shared-Memory Parallel Decoding (LSPD) to translate the compressed representation into Tensor-Core-ready computation. $B_t$ and $\widetilde{B}_t$ denote the logical union masks before and after row reordering, respectively. $P_t=\|B_t\|_0$ and $\widetilde{P}_t=\|\widetilde{B}_t\|_0$ denote their numbers of active entries, and $V_t^{(q)}$ denotes the compressed value stream of Area $q$.}
    \label{fig:overview}
\end{figure*}

Figure~\ref{fig:overview} shows UnionSparse's co-design of \emph{representation optimization} for lower metadata overhead and \emph{execution optimization} for efficient Tensor Core use.

The representation optimization part is shown on the left side of Fig.~\ref{fig:overview}. Starting from pruned low-bit quantized weights, UnionSparse first applies bitmap-aware row reordering within each local tile (LT) to reduce the number of active entries in the logical union mask. This step preserves the surviving nonzero values while reordering their rows to activate fewer union-mask positions. The reordered tile is then encoded using Index-Efficient Bitmap Encoding (IE-BME), in which four quantized payload areas share a single logical union bitmap, and their payloads are stored in aligned compressed streams. This design reduces index-heavy metadata while producing a regularized hierarchical layout suitable for later staging and decoding.

The execution optimization part is shown on the right side of Fig.~\ref{fig:overview}. During kernel execution, UnionSparse adopts a Decoupled Warp-Specialized Producer-Consumer (DW-PC) pipeline to overlap tile prefetch with current-tile computation. A producer warp prefetches the next weight and activation tiles into shared memory, while the consumer warps process the current tile. This decoupling enables overlap between next-tile data movement and current-tile computation while reducing concurrent warp-level pressure on the cache hierarchy. Once the compressed weight tile is staged in shared memory, Low-Bit Shared-Memory Parallel Decoding (LSPD) reconstructs the Tensor-Core-ready weight fragment directly into registers. In parallel, the corresponding activation tile is loaded into registers without sparse decoding. The decoded weight fragment and the activation fragment are then consumed by Tensor Core MMA instructions to produce the output tile.

IE-BME and bitmap-aware row reordering reduce union-mask activity and metadata, while DW-PC and LSPD exploit the regular layout for efficient staging, decoding, and computation.

\section{Representation Optimization}
    \subsection{IE-BME Representation}
\begin{figure*}[t]
    \centering
    \includegraphics[width=0.83\textwidth]{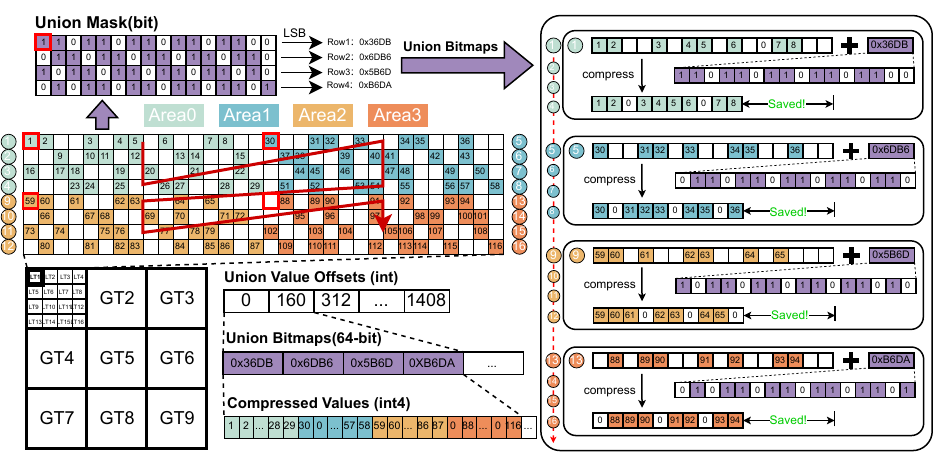}
\caption{Hierarchical IE-BME representation. A sparse weight matrix is partitioned into global tiles (GTs), each containing multiple local tiles (LTs). For an LT $W_t$, IE-BME constructs a shared logical union mask $B_t$, LT-level offsets $o_t$, and $Q$ aligned compressed payload streams $V_t^{(q)}$, where $q\in\{0,1,\ldots,Q-1\}$. LT is actually $16\times 64$, shown as $8\times 32$ for illustration.}

    \label{fig:encode}
\end{figure*}

IE-BME is designed to improve payload-to-metadata efficiency in low-bit sparse LLM inference while preserving a layout that can be decoded efficiently during online execution. As shown in Fig.~\ref{fig:encode}, the key idea of IE-BME is to amortize a single logical bitmap across multiple low-bit payload regions, so that a single index structure can describe several compressed values at once. At the same time, the resulting payload layout is regularized to support shared-memory staging and direct fragment generation in the later kernel. To make the encoding process easier to follow, Fig.~\ref{fig:encode} should be read from left to right: IE-BME first groups area-aligned values into logical positions, then constructs a shared union mask for these groups, and finally materializes the encoded LT as three storage arrays, namely Union Value Offsets, Union Bitmaps, and Compressed Values.

\textbf{Area Partitioning.}
Consider a quantized local tile $W_t \in \mathcal{Q}_b^{H_L \times K_L}$, where $\mathcal{Q}_b$ denotes the alphabet of a signed $b$-bit quantizer. IE-BME partitions $W_t$ into an $R_A \times C_A$ grid of areas. The total number of areas and a size of each area is
\[
Q = R_A C_A, \qquad
\frac{H_L}{R_A} \times \frac{K_L}{C_A}
\]

Accordingly, the tile can be written as
\[
W_t =
\bigl[\, W_t^{(q)} \,\bigr]_{q=0}^{Q-1},
\qquad
W_t^{(q)} \in \mathcal{Q}_b^{\frac{H_L}{R_A}\times \frac{K_L}{C_A}} 
\]

Each spatial position $(i,j)$, with
\[
0 \le i < \frac{H_L}{R_A}, \qquad
0 \le j < \frac{K_L}{C_A}
\]
defines one logical grouped position shared by all areas. The corresponding area-aligned value group is
\[
\mathbf{g}_t(i,j)=
\bigl(W_t^{(0)}(i,j), W_t^{(1)}(i,j), \ldots, W_t^{(Q-1)}(i,j)\bigr).
\]

\textbf{Shared Union Mask.}
IE-BME assigns one logical activity bit to each grouped position $\mathbf{g}_t(i,j)$. Instead of storing $Q$ independent bitmaps, IE-BME associates this group with a single shared logical union mask:
\[
B_t(i,j)
=
\mathbb{I}\!\left(
\bigvee_{q=0}^{Q-1}\bigl(W_t^{(q)}(i,j)\neq 0\bigr)
\right),
\quad
B_t \in \{0,1\}^{\frac{H_L}{R_A}\times \frac{K_L}{C_A}} 
\]

Thus, $B_t(i,j)=1$ indicates that at least one value in $\mathbf{g}_t(i,j)$ is nonzero, while $B_t(i,j)=0$ indicates that all $Q$ values in this group are zero. One logical union-mask entry therefore represents the activity of $Q$ payload values at the same area-aligned position. This shared-mask construction is the key reason why IE-BME reduces metadata overhead compared with per-area bitmap encoding. In the UnionSparse illustration of Fig.~\ref{fig:encode}, we use a $2\times2$ partition, i.e., $R_A=C_A=2$ and $Q=4$, to visualize this process. For example, the red-framed logical position in Fig.~\ref{fig:encode} groups four values $[1,30,59,0]$ from Area0--Area3. Since at least one value is nonzero, IE-BME stores a single union-mask bit $1$ for this group. This active bit contributes one compact payload rank, and the four aligned area streams append $1$, $30$, $59$, and $0$, respectively. The last value is still written because the shared union bit is active. Keeping this area-specific zero preserves stream alignment so that the same compact rank can later index all four area streams. In contrast, if the grouped values are $[0,0,0,0]$, the union-mask bit is $0$, and no payload entry is emitted for any area.

\textbf{IE-BME Storage Layout.}
Let $P_t=\|B_t\|_0$ denote the number of active entries in the logical union mask of LT $t$:
\[
P_t=\sum_{i=0}^{\frac{H_L}{R_A}-1}
\sum_{j=0}^{\frac{K_L}{C_A}-1}
B_t(i,j).
\]
IE-BME materializes the encoded sparse matrix using the following three arrays.

\emph{Union Bitmaps.}
The logical mask $B_t$ is scanned in row-major order and packed into bitmap words. In the simplified example of Fig.~\ref{fig:encode}, the leftmost bit of each displayed mask row is treated as the least significant bit (LSB), so the four logical rows are packed as \texttt{0x36DB}, \texttt{0x6DB6}, \texttt{0x5B6D}, and \texttt{0xB6DA}. These packed words are concatenated across LTs in the global Union Bitmaps array. In the actual implementation, the same rule is applied to the full LT shape and stored using 64-bit bitmap words.

\emph{Union Value Offsets.}
The offset array records the starting position of each LT's compressed payload segment in the global Compressed Values array. Conceptually, the offset of LT $t$ is a prefix sum over the compressed payload lengths of preceding LTs:
\[
o_t=\sum_{\tau<t} QP_\tau .
\]
Therefore, the compressed payload segment of LT $t$ occupies the range $[o_t,\,o_t+QP_t)$ in the global value array. In the illustrated LT of Fig.~\ref{fig:encode}, each of the four logical mask rows contains ten active bits, so $P_t=40$. Since $Q=4$, this LT contributes $QP_t=160$ int4 payload values, which explains the next offset value $160$ in the Union Value Offsets array.

\emph{Compressed Values.}
For every active logical position, IE-BME assigns a compact rank according to the number of preceding active union-mask entries:
\[
\rho_t(i,j)=\sum_{(u,v)\prec(i,j)} B_t(u,v),
\]
where $(u,v)\prec(i,j)$ denotes row-major order over the logical mask. If $B_t(i,j)=1$, IE-BME appends one value from every area using the same compact rank:
\[
V_t^{(q)}[\rho_t(i,j)] = W_t^{(q)}(i,j),
\qquad q\in\{0,1,\ldots,Q-1\}.
\]
Equivalently, when the $Q$ aligned streams of LT $t$ are stored consecutively in the global Compressed Values array, the storage address is
\[
\mathrm{CV}[o_t+qP_t+\rho_t(i,j)] = W_t^{(q)}(i,j).
\]
Positions with $B_t(i,j)=0$ emit no values. Positions with $B_t(i,j)=1$ emit $Q$ aligned values even if some area-specific values are zero, as shown by the red-framed $[1,30,59,0]$ example. This aligned layout is the key reason why the later LSPD stage can use one bitmap traversal and one compact rank to gather all area streams, rather than maintaining separate sparse indices for each low-bit payload region. Consequently, IE-BME reduces metadata overhead while keeping the payload layout regular enough for shared-memory staging and direct fragment generation.

\subsection{Bitmap-Aware Row Reordering}

Although IE-BME reduces metadata overhead through shared union-mask encoding, its efficiency still depends on how many grouped positions in the logical union mask become active. For a fixed set of retained nonzeros, this number is not invariant: different row arrangements within a local tile may activate different numbers of union-mask entries. Therefore, beyond the sparse format itself, representation efficiency also depends on how rows are grouped across the $R_A$ row partitions of each LT. UnionSparse exploits this flexibility through an offline, bitmap-aware row-reordering step.

\textbf{Objective.}
Under a general $R_A \times C_A$ IE-BME partition, one local tile $W_t \in \mathcal{Q}_b^{H_L \times K_L}$ is decomposed into $Q=R_A C_A$ areas, each of size $\frac{H_L}{R_A}\times\frac{K_L}{C_A}$. Each logical union-mask entry therefore corresponds to one grouped position spanning all $R_A$ row partitions and $C_A$ column partitions. For a fixed local tile, the total number of active entries in the logical union mask is determined by how rows are assigned to the $R_A$ row partitions. The objective of row reordering is to find a row permutation that minimizes this number before compression.

\textbf{Grouped Cost Formulation.}
Let
\[
M_t = \mathbb{I}(W_t \neq 0) \in \{0,1\}^{H_L \times K_L}
\]
denote the binary nonzero mask of tile $t$. For a candidate row group
\[
\mathbf{r}=(r_0,r_1,\ldots,r_{R_A-1}),
\]
define the grouped activity at grouped column index $j$ as
\begin{equation}
\label{eq:grouped-activity}
\phi_t(\mathbf{r};j)
=
\bigvee_{a=0}^{R_A-1}\;\bigvee_{c=0}^{C_A-1}
M_t\!\left(r_a,\; j+c\frac{K_L}{C_A}\right),
\quad
0 \le j < \frac{K_L}{C_A}.
\end{equation}

The corresponding grouped cost is
\begin{equation}
\label{eq:grouped-cost}
c_t(\mathbf{r})
=
\sum_{j=0}^{\frac{K_L}{C_A}-1}\phi_t(\mathbf{r};j),
\end{equation}
which counts the number of logical union-mask entries activated by grouping the rows in $\mathbf{r}$. Hence, minimizing the total number of active mask entries is equivalent to partitioning the rows of $W_t$ into groups of size $R_A$ with minimum total grouped cost.

\textbf{Grouped Subset Dynamic Programming.}
To support arbitrary $R_A$ and $C_A$, UnionSparse solves the row-reordering problem with a subset dynamic programming (DP) procedure. Let $S \subseteq \{0,1,\ldots,H_L-1\}$ denote the set of rows that have not yet been grouped, with $|S|$ divisible by $R_A$. The DP state value $F(S)$ is defined as the minimum grouped cost achievable on the remaining rows in $S$. For each state, we select the smallest row index $r \in S$, enumerate all row groups $\mathbf{r}\subseteq S$ of size $R_A$ that contain $r$, and minimize
\begin{equation}
\label{eq:group-dp}
F(S)
=
\min_{\mathbf{r}\subseteq S,\; |\mathbf{r}|=R_A,\; r\in\mathbf{r}}
\left(
c_t(\mathbf{r}) + F(S\setminus \mathbf{r})
\right).
\end{equation}
The base case is $F(\varnothing)=0$.

After the optimal row groups are obtained, we canonicalize the result for deterministic output: rows inside each group are sorted in ascending order, and all groups are then sorted lexicographically. Let $r_{g,p}$ denote the $p$-th row in the $g$-th sorted row group, where
\[
g=0,1,\ldots,\frac{H_L}{R_A}-1,
\qquad
p=0,1,\ldots,R_A-1.
\]
Then the reordered permutation is constructed by concatenating rows partition-wise across all groups:
\[
\pi_t^\star
=
\bigl[\, r_{g,p}\ \big|\ p=0,\ldots,R_A-1,\; g=0,\ldots,\frac{H_L}{R_A}-1 \,\bigr].
\]

Algorithm~\ref{alg:group-dp-reorder} summarizes this solver. In our implementation, row reordering is performed offline and independently for each LT, so it does not introduce any online decoding or scheduling overhead.

\begin{algorithm}[t]
\caption{Bitmap-Aware Row Reordering via Grouped Subset DP}
\label{alg:group-dp-reorder}
\footnotesize
\begin{algorithmic}[1]
\Require Binary local-tile mask $M_t \in \{0,1\}^{H_L \times K_L}$, row partition count $R_A$, column partition count $C_A$
\Ensure Reordered permutation $\pi_t^\star$
\State Precompute grouped costs $c_t(\mathbf{r})$ for all row groups $\mathbf{r}$ of size $R_A$
\State Define DP state $F(S)$ over row subsets $S \subseteq \{0,1,\ldots,H_L-1\}$ with $|S| \equiv 0 \pmod{R_A}$
\State Initialize $F(\varnothing) \gets 0$
\ForAll{valid DP states $S$ in memoized recursion}
    \State Let $r$ be the smallest row index in $S$
    \ForAll{row groups $\mathbf{r}\subseteq S$ such that $|\mathbf{r}|=R_A$ and $r\in\mathbf{r}$}
        \State Update
        \[
        F(S) \gets \min \bigl(F(S),\; c_t(\mathbf{r}) + F(S\setminus\mathbf{r})\bigr)
        \]
    \EndFor
\EndFor
\State Recover the optimal row groups from the DP solution
\State Sort rows inside each group and sort all groups lexicographically
\State Construct $\pi_t^\star$ by concatenating rows partition-wise across all groups
\State \Return $\pi_t^\star$
\end{algorithmic}
\end{algorithm}

\textbf{Complexity and Offline Cost.} For each LT, bitmap-aware row reordering consists of two stages. First, it builds a grouped-cost table by enumerating all candidate row groups of size \(R_A\) and computing the number of union-mask entries activated by each group over the \(K_L/C_A\) grouped columns. This stage takes \(O(\binom{H_L}{R_A} R_A K_L)\) time. Second, the subset DP searches for the minimum-cost partition over row subsets. The number of DP states grows exponentially with \(H_L\). For each state, the smallest remaining row is fixed and only candidate groups containing this row are enumerated. Therefore, the number of DP transitions is upper bounded by \(O(2^{H_L}\binom{H_L-1}{R_A-1})\). Overall, the per-LT time complexity is \(O\!\left(\binom{H_L}{R_A} R_A K_L + 2^{H_L}\binom{H_L-1}{R_A-1}\right)\), and the space complexity is \(O(2^{H_L}+\binom{H_L}{R_A})\). For a weight matrix of size \(M \times K\), the number of LTs is \((M/H_L)(K/K_L)\), so the matrix-level cost is the above per-LT cost multiplied by the number of LTs. For a full model, the cost is summed over all weight matrices. This step is performed once during offline weight-format generation, and the reordered IE-BME weights are cached for subsequent online inference, introducing no online decoding or scheduling overhead. In our measurement, running this row-reordering step on OPT-13B takes about 4.07 hours using 14 threads on the Arm Neoverse-V3AE CPU of Jetson AGX Thor.

\textbf{Practical Effect.}
Bitmap-aware row reordering does not change the number or values of surviving nonzeros, it only changes their row arrangement inside each LT. Its benefit arises from reducing the number of active grouped positions, which in turn decreases the size of the logical union mask and shortens the aligned area streams produced by IE-BME. Consequently, the representation becomes more payload-efficient before entering the online kernel stage.

\section{Execution Optimization}

    \subsection{Decoupled Warp-Specialized Producer-Consumer Pipeline}

\begin{figure}[t]
    \centering
    \includegraphics[width=0.85\linewidth]{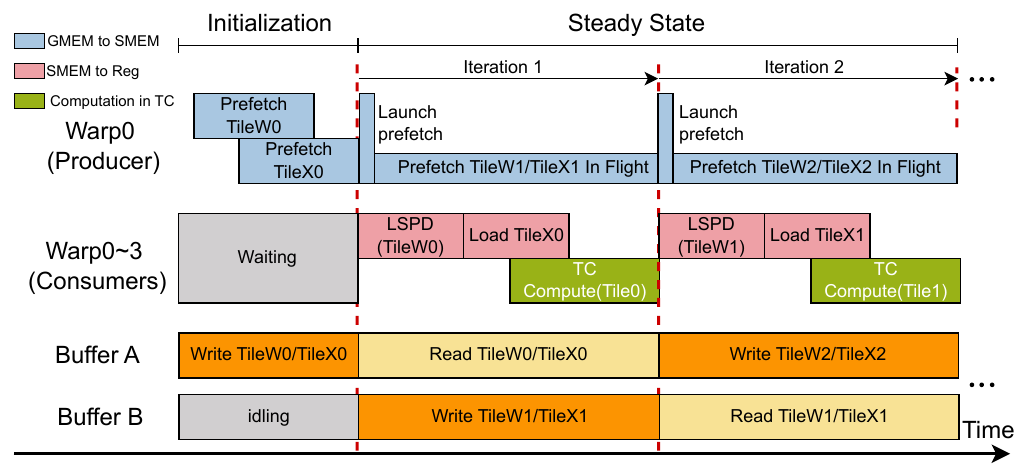}
    \caption{Decoupled Warp-Specialized Producer-Consumer (DW-PC) pipeline. Next-tile prefetch is decoupled from current-tile processing at the warp level. The producer warp prefetches the next weight and activation tiles, while the consumer warps perform LSPD, activation loading, and Tensor Core computation on the current tile. This double-buffered design overlaps data movement with computation and reduces concurrent pressure on the memory hierarchy.}
    \label{fig:pipeline}
\end{figure}

Figure~\ref{fig:pipeline} illustrates the Decoupled Warp-Specialized Producer-Consumer (DW-PC) pipeline used in the UnionSparse kernel. The key motivation is that, under small-batch decoding, the computation associated with one tile is often insufficient to hide memory latency if all warps synchronously participate in loading the next tile. In addition, concurrent requests issued by multiple warps can aggravate pressure on the cache hierarchy. UnionSparse addresses these issues by decoupling next-tile prefetch from current-tile processing at the warp level.

\textbf{Double Buffering Strategy.}
UnionSparse maintains two shared-memory buffers for the weight-tile stream and two for the activation-tile stream. At each iteration, one buffer pair serves as the read buffer for the current tiles, while the other serves as the write buffer for the next tiles. During initialization, the first tile pair, namely $\mathrm{TileW}_0$ and $\mathrm{TileX}_0$, is staged into the read buffers. After the current iteration finishes, execution reaches a buffer-swap barrier, at which the write buffers become the read buffers for the next iteration. This ping-pong organization enables next-tile staging and current-tile processing to proceed in an overlapped manner.

 \textbf{Warp-Specialized Roles.}
As shown in Fig.~\ref{fig:pipeline}, Warp~0 is designated as the \emph{producer warp}. Its role is to issue the global-to-shared-memory prefetch for the next tiles, i.e., $\mathrm{TileW}_{k+1}$ and $\mathrm{TileX}_{k+1}$. The remaining warps are not involved in this next-tile prefetch. Instead, they continue processing the current tiles already staged in shared memory. Importantly, Warp~0 is not permanently detached from computation. Once the prefetch has been launched, it rejoins the other warps as a \emph{consumer} and participates in current-tile execution. Therefore, the effective consumer set consists of Warp~0--3.

 \textbf{Steady-State Schedule.}
In the steady state, the producer warp launches the prefetch of $\mathrm{TileW}_{k+1}$ and $\mathrm{TileX}_{k+1}$ into the write buffers, while the consumer warps process $\mathrm{TileW}_k$ and $\mathrm{TileX}_k$ from the read buffers. Current-tile processing consists of three stages: LSPD on $\mathrm{TileW}_k$, loading $\mathrm{TileX}_k$ from shared memory to registers, and Tensor Core MMA on the resulting fragments. These two branches are synchronized only at the buffer-switching boundary. Once the prefetched next tiles are ready in shared memory, execution advances to the next iteration.

    \subsection{Low-Bit Shared-Memory Parallel Decoding}

\begin{figure*}[t]
    \centering
    \includegraphics[width=0.87\textwidth]{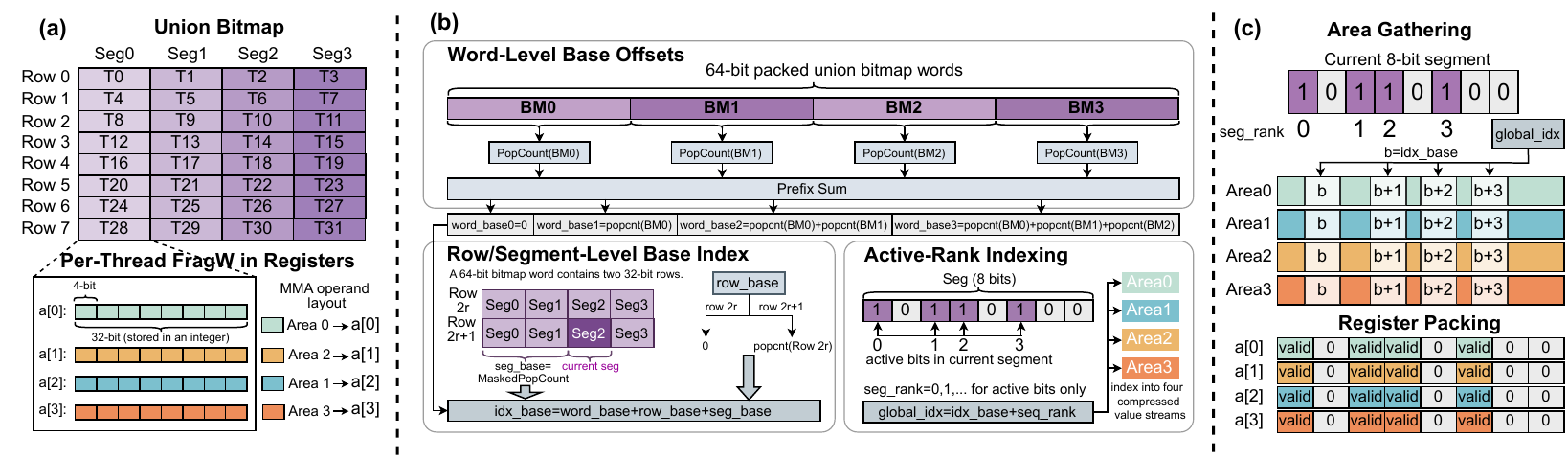}
    \caption{Low-Bit Shared-Memory Parallel Decoding (LSPD). (a) Thread-to-segment mapping on the LT-level union bitmap. (b) Bitmap-guided offset generation using packed union-bitmap words. (c) Gathering aligned area streams and packing them into registers, enabling direct reconstruction of MMA-ready fragments from the compressed IE-BME representation.}
    \label{fig:lspd}
\end{figure*}

The compressed weight tile $\mathrm{TileW}_k$ staged in shared memory cannot be directly consumed by Tensor Core MMA. Instead, it must be transformed into warp-level register fragments that match the operand layout required by the MMA instruction. UnionSparse performs this transformation using \emph{Low-Bit Shared-Memory Parallel Decoding (LSPD)}. Figure~\ref{fig:lspd} illustrates this procedure using instantiation $R_A=C_A=2$ for example.

 \textbf{Thread-to-Segment Mapping.}
Under a general $R_A \times C_A$ IE-BME partition, the logical union mask of one LT has size
\[
H_B \times K_B
\;=\;
\frac{H_L}{R_A}\times \frac{K_L}{C_A}.
\]

For decoding, each logical row is further partitioned into equal-width bitmap segments. Let the segment width be $G$ bits, and let
\[
S = \frac{K_B}{G}
\]
denote the number of segments per logical row. Then one thread is assigned to one bitmap segment, and one warp processes multiple logical rows in parallel. Figure~\ref{fig:lspd}(a) shows this thread-to-segment mapping under the current implementation.

 \textbf{Bitmap-Guided Offset Generation.}
Let
\[
B_t \in \{0,1\}^{H_B \times K_B}
\]
be the logical union mask of tile $t$. For a thread assigned to row $r$ and segment $g$, define the starting column of the segment as
\[
j_g = gG.
\]

The starting payload index of this segment can be written in a parameterized form as
\[
\mathrm{idx\_base}(r,g)
=
\sum_{u=0}^{r-1}\sum_{v=0}^{K_B-1} B_t(u,v)
+
\sum_{v=0}^{j_g-1} B_t(r,v).
\]

For each active bit inside the current segment, the final access index is
\[
\mathrm{global\_idx}
=
\mathrm{idx\_base}(r,g)+\mathrm{seg\_rank},
\]
where $\mathrm{seg\_rank}$ enumerates only the active positions within the segment. Figure~\ref{fig:lspd}(b) shows the corresponding offset-generation flow.

 \textbf{Area Gathering and Register Packing.}
Once $\mathrm{global\_idx}$ is obtained, the same index is applied to all $Q=R_A C_A$ compressed area streams:
\[
V_t^{(q)}[\mathrm{global\_idx}],
\qquad q\in\{0,1,\ldots,Q-1\}.
\]

Hence, one active union-mask entry selects a group of $Q$ low-bit payload values, one from each area stream. Inactive positions do not trigger payload fetches and are padded with zeros instead. For example, Fig.~\ref{fig:lspd}(c) shows an 8-bit segment \texttt{10110100}. Its active bits are at positions 0, 2, 3, and 5. Given the segment base $b=\mathrm{idx\_base}$, LSPD maps them to consecutive payload indices $b$, $b+1$, $b+2$, and $b+3$, which are reused across the four aligned area streams under $R_A=C_A=2$. The zero bits do not fetch payloads and are filled with zeros in the per-thread register fragment. The gathered values are then packed into a set of per-area register subvectors that serve as the decoded weight fragment for Tensor Core computation.

 \textbf{Design Implication.}
LSPD is important for two reasons. First, it eliminates explicit sparse-coordinate storage and replaces it with bitmap-guided online index reconstruction. Second, it maps the recovered low-bit values directly into the register layout required by Tensor Core MMA. Together, these properties reduce online decoding overhead and enable the compact IE-BME representation to be consumed efficiently by the UnionSparse kernel.

    \subsection{UnionSparse Kernel}

\begin{algorithm}[t]
\caption{High-Performance UnionSparse Kernel}
\label{alg:unionsparse-kernel}
\footnotesize
\begin{algorithmic}[1]
\Require Compressed sparse matrix $W$ in IE-BME format, dense activation matrix $X$, SplitK
\Ensure Matrix $Y$
\State $\mathrm{BatchID} \gets \lfloor \mathrm{blockIdx.y}/(M/T_M) \rfloor$
\State $\mathrm{TileY} \gets \mathrm{blockIdx.y} \bmod (M/T_M),\quad \mathrm{TileX} \gets \mathrm{blockIdx.x}$
\State $(\mathrm{NumIter}, \mathrm{KBlockStart}) \gets \Call{InitSchedule}{\mathrm{BatchID}, \mathrm{SplitK}}$
\State $\mathrm{is\_producer} \gets (\lfloor \mathrm{threadIdx.x}/32 \rfloor = 0)$
\State \Call{InitDoubleBuffers}{}
\State Initialize register fragment $\mathrm{FragY}$
\State \textcolor{blue}{// Initialization}
\State \textsc{PrefetchTilePair}$(W, X, \mathrm{TileY}, \mathrm{TileX},$
\Statex \hspace{2em}$\mathrm{KBlockStart}, \mathrm{ReadBuffer})$
\State \texttt{cp.async.commit()}
\If{$\mathrm{is\_producer}$}
    \State \texttt{cp.async.wait\_group(0)} \algcmt{$\mathrm{TileW}_0$, $\mathrm{TileX}_0$ ready}
\EndIf
\State \texttt{\_\_syncthreads()} \algcmt{Tile-0-ready barrier}
\State \textcolor{blue}{// Steady-state loop}
\For{$k = 0$ \textbf{to} $\mathrm{NumIter}-2$}
    \State \textsc{ProducerPrefetchNext}$(W, X, \mathrm{TileY}, \mathrm{TileX}, \mathrm{KBlockStart}+k+1, \mathrm{WriteBuffer})$
    \State \texttt{cp.async.commit()} \algcmt{launch next-tile prefetch}
    \State $\mathrm{FragY} \gets \Call{ProcessCurrentTile}{\mathrm{ReadBuffer}, \mathrm{FragY}}$
    \If{$\mathrm{is\_producer}$}
        \State \texttt{cp.async.wait\_group(0)} \algcmt{next tile pair ready}
    \EndIf
    \State \texttt{\_\_syncthreads()} \algcmt{buffer-swap barrier}
    \State \Call{SwapBuffers}{ReadBuffer, WriteBuffer}
\EndFor
\State \textcolor{blue}{// Epilogue}
\State $\mathrm{FragY} \gets \Call{ProcessCurrentTile}{\mathrm{ReadBuffer}, \mathrm{FragY}}$
\State \Call{StoreTile}{$\mathrm{FragY}, \mathrm{TileY}, \mathrm{TileX}, \mathrm{BatchID}, \mathrm{ReductionWorkspace}$}
\State $Y \gets \Call{Reduction}{\mathrm{ReductionWorkspace}}$
\end{algorithmic}
\end{algorithm}

Algorithm~\ref{alg:unionsparse-kernel} summarizes the execution flow of the UnionSparse kernel. The kernel is organized as a double-buffered producer-consumer pipeline, in which Warp~0 prefetches the next tile pair and Warp~0--3 process the current tile pair. This organization matches the DW-PC schedule in Fig.~\ref{fig:pipeline}, while 
LSPD instantiates the low-bit sparse decoding step inside \textsc{ProcessCurrentTile}, as illustrated in Fig.~\ref{fig:lspd}.

The initialization stage is given by lines 8--13. The first weight tile and activation tile are prefetched into the read buffer in line 8 and committed in line 9. The producer-side \texttt{cp.async.wait\_group(0)} in line 11 confirms that the first tile pair is ready, and the block-wide \texttt{\_\_syncthreads()} in line 13 establishes the tile-ready barrier. This stage prepares the pipeline so that the subsequent iterations can immediately start from a valid tile pair in shared memory.

\begin{figure*}[t]
  \centering
  \includegraphics[width=0.87\textwidth]{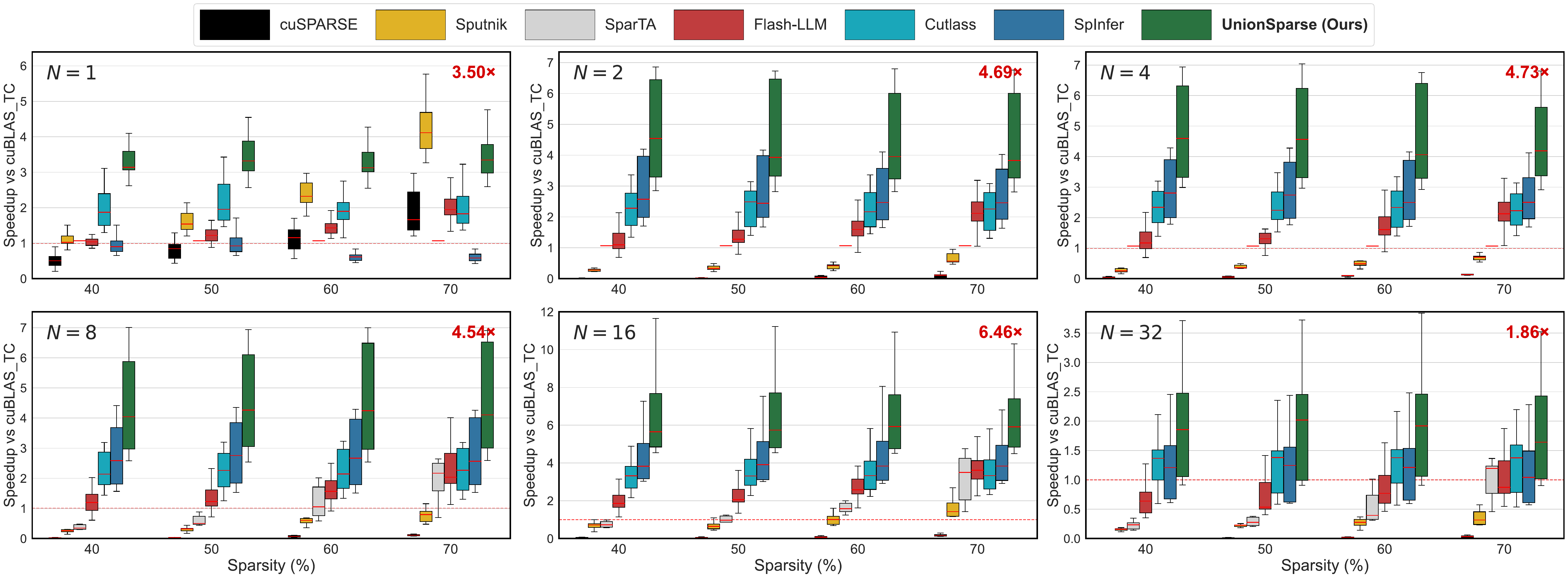}
  \caption{Kernel-Level Speedup across Decode Widths $N$ (Batch Sizes) and Sparsity under W4A4 Quantization. Each subplot corresponds to a different decode width (batch size), with sparsity varying along the x-axis.}
  \label{fig:int4-speedup-by-n}
\end{figure*}

The steady-state loop is implemented in lines 15--23. In line 16, the producer warp launches the prefetch of the next tile pair into the write buffer, and line 17 commits the corresponding asynchronous copy group. Meanwhile, the current tile pair in the read buffer is processed in line 18 through \textsc{ProcessCurrentTile}, which fuses LSPD decoding, activation loading, and Tensor Core MMA into one logical step. After the producer confirms that the next tile pair is ready in line 20, the block reaches the buffer-swap barrier in line 22 and exchanges the read and write buffers in line 23. This loop therefore realizes the intended overlap between next-tile staging and current-tile computation.

The epilogue is given by lines 26--28. After the steady-state loop terminates, one tile pair remains in the read buffer. The kernel processes this final tile pair, stores the partial output tile, and then performs reduction to obtain the final matrix $Y$. In this way, UnionSparse executes compressed sparse tiles through one unified pipeline that combines asynchronous staging and low-bit sparse computation.

\section{Evaluation}

\subsection{Evaluation Setup}

\textbf{Platform.}
All experiments are conducted on a Jetson AGX Orin 64GB edge platform\cite{nvidia_jetson_orin} (Ampere GPU, Compute Capability 8.7) running Ubuntu 22.04.5 LTS, JetPack 6.1, and CUDA 12.6, with MAX-N mode enabled. The code is compiled using GCC 11.4.0 and NVCC 12.6.

\textbf{Kernel Benchmark Scope.}
We adopt a representative W4A4 quantization setting for evaluation, as it reflects a typical low-bit sparse LLM inference scenario on edge GPUs. For kernel-level evaluation, we construct a benchmark corpus from representative weight-matrix shapes extracted from LLaMA2-7B, LLaMA2-13B\cite{touvron2023llama2}, OPT-13B\cite{opt}, Qwen2-7B\cite{qwen}, and LLaMA3-8B\cite{touvron2024llama3}. The corpus covers decode widths $N\in\{1,2,4,8,16,32\}$ and sparsity levels of 40\%, 50\%, 60\%, and 70\%, resulting in 832 valid benchmark cases. Unless otherwise stated, each kernel is executed 25 times, and the average latency is reported. Let $c$ denote one benchmark case, including the matrix shape, decode width $N$, and sparsity level. 
We use $T_M(c)$ to denote the average latency of method $M$ on case $c$. Nsight Compute is used for profiling and metric analysis.

\textbf{Baselines and Fairness.}
For kernel-level comparison, we evaluate UnionSparse against SpInfer~\cite{spinfer}, FlashLLM~\cite{flashllm}, CUTLASS~\cite{nvidia_cutlass}, cuBLAS~\cite{nvidia_cublas}, cuSPARSE~\cite{nvidia_cusparse}, Sputnik~\cite{gale2020sparse}, and SparTA~\cite{zheng2022sparta}. These baselines cover both dense and sparse implementations, as well as Tensor-Core-based and CUDA-core-based execution. Since these baselines do not all support an identical W4A4 sparse Tensor-Core path, each method is evaluated under its native or supported execution configuration on the target platform. Dense baselines serve as Tensor-Core GEMM references without exploiting sparsity, while cuSPARSE, Sputnik, and SparTA use their native CUDA-core sparse paths. For LLM-oriented sparse baselines, we extend SpInfer to INT4 Tensor-Core execution while preserving its original representation and scheduling, and evaluate FlashLLM with its original load-sparse, compute-dense FP16 path using inputs pre-converted to the required format. All reported kernel latencies exclude offline preprocessing, including pruning, quantization, format conversion, and reordering. They measure only the online execution latency of each method. Kernel-level results are reported as latency and speedup relative to cuBLAS-TC unless otherwise stated.

\textbf{End-to-End Setting.}
For end-to-end evaluation, we integrate UnionSparse into FasterTransformer~\cite{fastertransformer} under W4A4 deployment and evaluate OPT-13B~\cite{opt} on the same platform. We use Wanda~\cite{wanda} for pruning and Atom~\cite{zhao2024atom} for W4A4 quantization.

\textbf{Energy Measurement.}
We further measure end-to-end inference energy on Jetson AGX Orin using the on-board INA3221 power sensor through the hwmon/sysfs interface, with a sampling interval of 50 ms. The idle power measured before the experiment is 3.19 W, which is subtracted from the integrated power trace to obtain incremental energy. The energy experiment uses OPT-13B under W4A4 quantization with input length 64 and output length 128. The Orin board runs in MAXN power mode, with CPU, GPU, and EMC frequencies locked to their maximum values.

\begin{figure*}[t]
  \centering
  \includegraphics[width=0.87\textwidth]{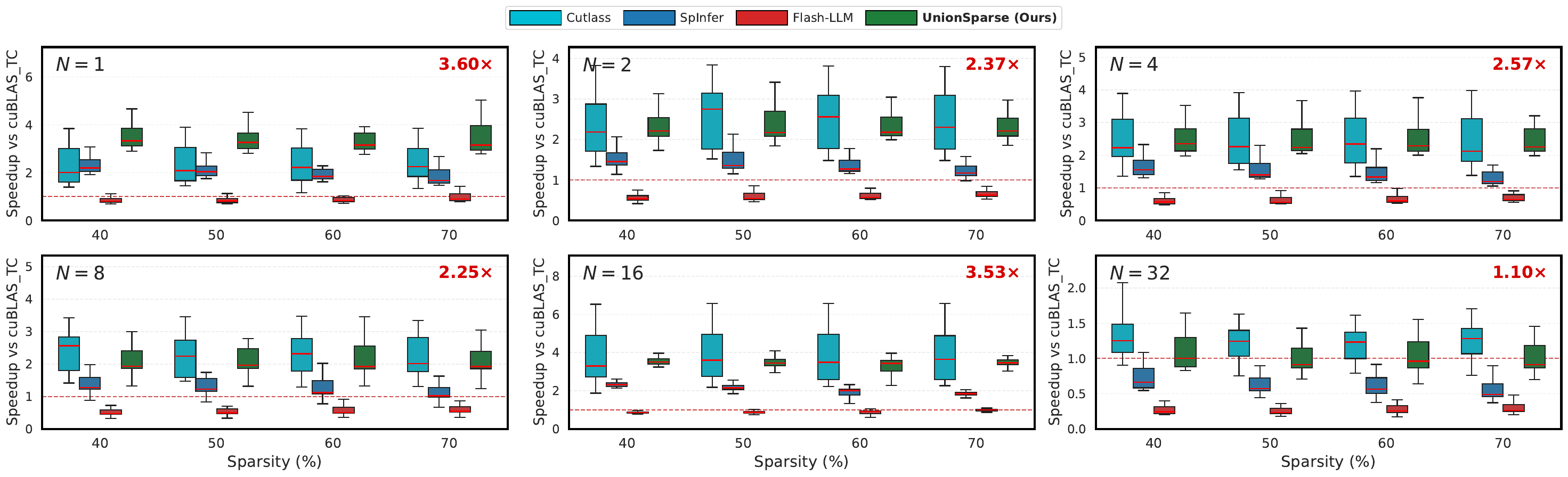}
  \caption{Kernel-Level Speedup across Decode Widths $N$ (Batch Sizes) and Sparsity under W2A4 Quantization. Each subplot corresponds to a different decode width (batch size), with sparsity varying along the x-axis.}
  \label{fig:w2a4-speedup-by-n}
\end{figure*}

\begin{table}[t]
\centering
\caption{Baseline characteristics in the evaluation.}
\label{tab:baseline_characteristics}
\footnotesize
\resizebox{\columnwidth}{!}{%
\begin{tabular}{lcccl}
\toprule
Method & Type & Backend & Scope & Notes \\
\midrule
cuBLAS-TC\cite{nvidia_cublas}& Dense  & Tensor Core & Kernel & Dense low-bit baseline \\
CUTLASS\cite{nvidia_cutlass}     & Dense  & Tensor Core & Kernel & Template dense kernel baseline \\
cuSPARSE\cite{nvidia_cusparse}    & Sparse & CUDA Core   & Kernel & Generic sparse baseline \\
Sputnik\cite{gale2020sparse}     & Sparse & CUDA Core   & Kernel & Sparse DNN baseline \\
SparTA\cite{zheng2022sparta}      & Sparse & CUDA Core   & Kernel & Hybrid sparse baseline \\
FlashLLM\cite{flashllm}    & Sparse & Tensor Core & Kernel/E2E & Load-sparse, compute-dense \\
SpInfer\cite{spinfer}     & Sparse & Tensor Core & Kernel/E2E & TCA-BME-based sparse baseline \\
UnionSparse & Sparse & Tensor Core & Kernel/E2E & IE-BME + DW-PC + LSPD \\
\bottomrule
\end{tabular}
}
\end{table}

\vspace{-10pt}
\subsection{Kernel Performance Comparison}

Figure~\ref{fig:int4-speedup-by-n} reports the kernel-level speedup of UnionSparse over multiple baselines across decode widths $N$ and sparsity levels under W4A4 quantization. Each subplot corresponds to one decode width, while the x-axis sweeps sparsity and the boxplots summarize the distribution across the benchmark corpus. Several trends are immediately visible. First, UnionSparse consistently forms the upper envelope across most workloads, indicating that its advantage is robust across model shapes rather than concentrated on a few outliers. Second, the advantage is observed against both dense Tensor-Core baselines (CUTLASS and cuBLAS-TC) and sparse baselines (SpInfer, FlashLLM, cuSPARSE, Sputnik, and SparTA). Third, the gain remains stable across sparsity levels from 40\% to 70\%, showing that the benefit of UnionSparse is not tied to a narrow sparsity regime.

To summarize these results more compactly, Table~\ref{tab:speedup_orin_thor_by_n} reports the arithmetic mean of per-case speedups of UnionSparse over each baseline at each decode width. Specifically, for a baseline $B$ and decode width $N$, let $\mathcal{C}_{B,N}$ be the set of benchmark cases with decode width $N$ for which both UnionSparse and baseline $B$ are supported. For each case $c\in\mathcal{C}_{B,N}$, we first compute the per-case speedup as $S_B(c)=T_B(c)/T_{\mathrm{UnionSparse}}(c)$. The value reported in Table II is then $\overline{S}_{B,N}=|\mathcal{C}_{B,N}|^{-1}\sum_{c\in\mathcal{C}_{B,N}} S_B(c)$. Unsupported baseline configurations are marked with ``--'' and excluded from the corresponding average.

Over the full W4A4 benchmark corpus in Jetson AGX Orin, UnionSparse achieves 1.56$\times$ speedup over CUTLASS, 3.46$\times$ over cuBLAS-TC, 2.30$\times$ over FlashLLM, and 1.43$\times$ over SpInfer. Compared with CUDA-core-based sparse baselines, the advantage is even larger, reaching 11.84$\times$ over SparTA (available for $N=8/16/32$), 19.63$\times$ over Sputnik, and 207.21$\times$ over cuSPARSE. These results indicate that conventional sparse formats become increasingly inefficient in the targeted low-bit sparse decoding regime, whereas UnionSparse preserves the benefits of sparse execution on Tensor Core.

\begin{table}[t]

\centering
\caption{Arithmetic-mean kernel speedup under W4A4 across decode widths $N$ on Orin/Thor.}
\label{tab:speedup_orin_thor_by_n}
\scriptsize
\setlength{\tabcolsep}{2.5pt}
\resizebox{\columnwidth}{!}{%
\begin{tabular}{lcccccc}
\toprule
\textbf{Baseline} & \textbf{$N=1$} & \textbf{$N=2$} & \textbf{$N=4$} & \textbf{$N=8$} & \textbf{$N=16$} & \textbf{$N=32$} \\
\midrule
SpInfer
& 4.60/1.09$\times$
& 1.66/1.09$\times$
& 1.66/1.09$\times$
& 1.63/1.53$\times$
& 1.53/1.18$\times$
& 1.55/1.19$\times$ \\

FlashLLM
& 2.57/1.36$\times$
& 3.08/1.35$\times$
& 3.07/1.35$\times$
& 2.97/1.34$\times$
& 2.56/0.99$\times$
& 2.39/0.61$\times$ \\

CUTLASS
& 1.81/---
& 2.10/---
& 2.08/---
& 2.03/---
& 1.90/---
& 1.45/--- \\

cuBLAS
& 3.50/1.29$\times$
& 4.69/1.30$\times$
& 4.73/1.33$\times$
& 4.54/1.43$\times$
& 6.46/1.97$\times$
& 1.86/0.37$\times$ \\

Sputnik
& 3.88/1.56$\times$
& 27.32/2.67$\times$
& 26.28/2.67$\times$
& 25.60/2.65$\times$
& 18.20/1.86$\times$
& 16.50/1.49$\times$ \\

cuSPARSE
& 9.38/1.87$\times$
& 219.83/73.93$\times$
& 225.31/74.74$\times$
& 235.53/77.53$\times$
& 235.22/106.62$\times$
& 318.02/135.37$\times$ \\

SparTA
& ---
& ---
& ---
& 13.89/---
& 12.14/---
& 9.50/--- \\
\bottomrule
\end{tabular}
}
\end{table}


A clear decode-width trend is also evident. UnionSparse is most effective in decode-oriented small- and medium-$N$ regimes, achieving 3.50$\times$, 4.69$\times$, 4.73$\times$, and 4.54$\times$ speedup over cuBLAS-TC at $N=1,2,4,$ and $8$, respectively. The gain remains high at $N=16$, and then narrows at $N=32$, where dense Tensor-Core kernels benefit more from regular dataflow and higher arithmetic intensity. 

We also conduct a kernel-level analysis under the more aggressive W2A4
quantization setting. As shown in Figure~\ref{fig:w2a4-speedup-by-n}, UnionSparse
remains competitive across decode widths and sparsity levels, especially in
small- and medium-$N$ regimes where sparse weight access and metadata handling
dominate the latency. The benefit becomes smaller at $N=32$, where the workload
is more compute-intensive. This confirms that UnionSparse's advantage is not
restricted to W4A4, but can also extend to lower-bit W2A4 inference.

Additionally, we evaluate UnionSparse on Jetson Thor GPU shown in Figure \ref{tab:speedup_orin_thor_by_n}, an SM110-based Blackwell
platform. CUTLASS is omitted because it does not provide a directly applicable SM110 baseline. SparTA is also excluded, as it only produces valid latency results
for $N \ge 8$ and thus does not cover the small-$N$ decode regimes considered
in this study. In this experiment, we don't use optimization method based on hardware characteristic of Jetson thor and choose RA=2 and CA=2, so the performance is not as good as that in Orin.

This trend is consistent with the motivation of UnionSparse: its main advantage comes from reducing metadata overhead and improving the efficiency of low-bit sparse execution, which is most critical when the workload remains bandwidth-sensitive and metadata cost is difficult to amortize.

\subsection{Ablation Study}

To quantify the contribution of each major component, we conduct a controlled kernel-level ablation study under a unified setting, where all configurations share the same matrix shape, sparsity ratio, quantization scheme, hardware platform, and decode scenario. Table~\ref{tab:kernel_ablation} reports the latency and representative profiling metrics of five configurations: the baseline without the proposed optimizations, \textit{LSPD only}, \textit{DB only}, \textit{LSPD+DB}, and \textit{LSPD+DB+Reorder}.

Three observations can be drawn. First, \textbf{LSPD provides the dominant standalone gain}, reducing latency from 126.40\,\textmu s to 102.72\,\textmu s, which shows that direct bitmap-guided fragment reconstruction is the most effective optimization in removing sparse decoding overhead. Second, \textbf{DB alone brings a smaller but complementary benefit}, reducing latency to 122.91\,\textmu s and improving scheduling-related metrics, indicating better overlap between memory access and computation. Third, \textbf{LSPD and DB are complementary}: combining them further reduces latency to 99.90\,\textmu s, outperforming either optimization in isolation.

Adding \textbf{Reorder} on top of LSPD+DB further reduces latency to 99.52\,\textmu s and yields small but consistent improvements in bandwidth utilization, issue efficiency, and Tensor Core activity. Although its incremental kernel-level gain is smaller than that of LSPD or DB, it confirms that the representation-side optimization also translates into measurable online benefit. 

Besides latency reduction, the auxiliary metrics in Table~\ref{tab:kernel_ablation} show a consistent trend: LSPD and DB improve bandwidth utilization and issue efficiency while reducing warp cycles per instruction, and Reorder further provides small but consistent gains in bandwidth and Tensor Core activity. These trends are consistent with the intended roles of the three optimizations.


%


Overall, the ablation results validate the two-part design of UnionSparse: execution-side optimizations contribute the main online speedup, while representation-side optimization provides an additional gain by improving the compactness and regularity of the compressed layout.

\begin{table}[t]
\centering
\caption{Kernel-level ablation study of LSPD, Double Buffer Pipeline, and reorder.}
\label{tab:kernel_ablation}
\footnotesize
\setlength{\tabcolsep}{3pt}
\renewcommand{\arraystretch}{1.05}
\begin{tabular}{ccc c c c c c c}
\toprule
\multicolumn{3}{c}{Optimization} &
Dur.$\downarrow$ &
BW$\uparrow$ &
Issue$\uparrow$ &
Warp Cyc$\downarrow$ &
TC$\uparrow$ &
ALU$\downarrow$ \\
LSPD & DB & Reorder &
(us) & (\%) & (\%) & /Inst & (\%) & (\%) \\
\midrule
 &  &  & 126.40 & 31.93 & 65.63 & 10.74 & 1.29 & 72.66 \\
\midrule
\checkmark &  &  & 102.72 & 33.12 & 67.48 & 9.98 & 1.56 & 68.77 \\
 & \checkmark &  & 122.91 & 32.70 & 65.97 & 9.44 & 1.30 & 70.61 \\
\midrule
\checkmark & \checkmark &  & 99.90 & 33.67 & 67.94 & 9.42 & 1.65 & 67.92 \\
\checkmark & \checkmark & \checkmark & 99.52 & 34.27 & 68.39 & 9.42 & 1.68 & 67.90 \\
\bottomrule
\end{tabular}
\end{table}




\vspace{-10pt}
\subsection{Profiling and Metric Analysis}

\begin{figure}[t]
  \centering
  \includegraphics[width=0.85\linewidth]{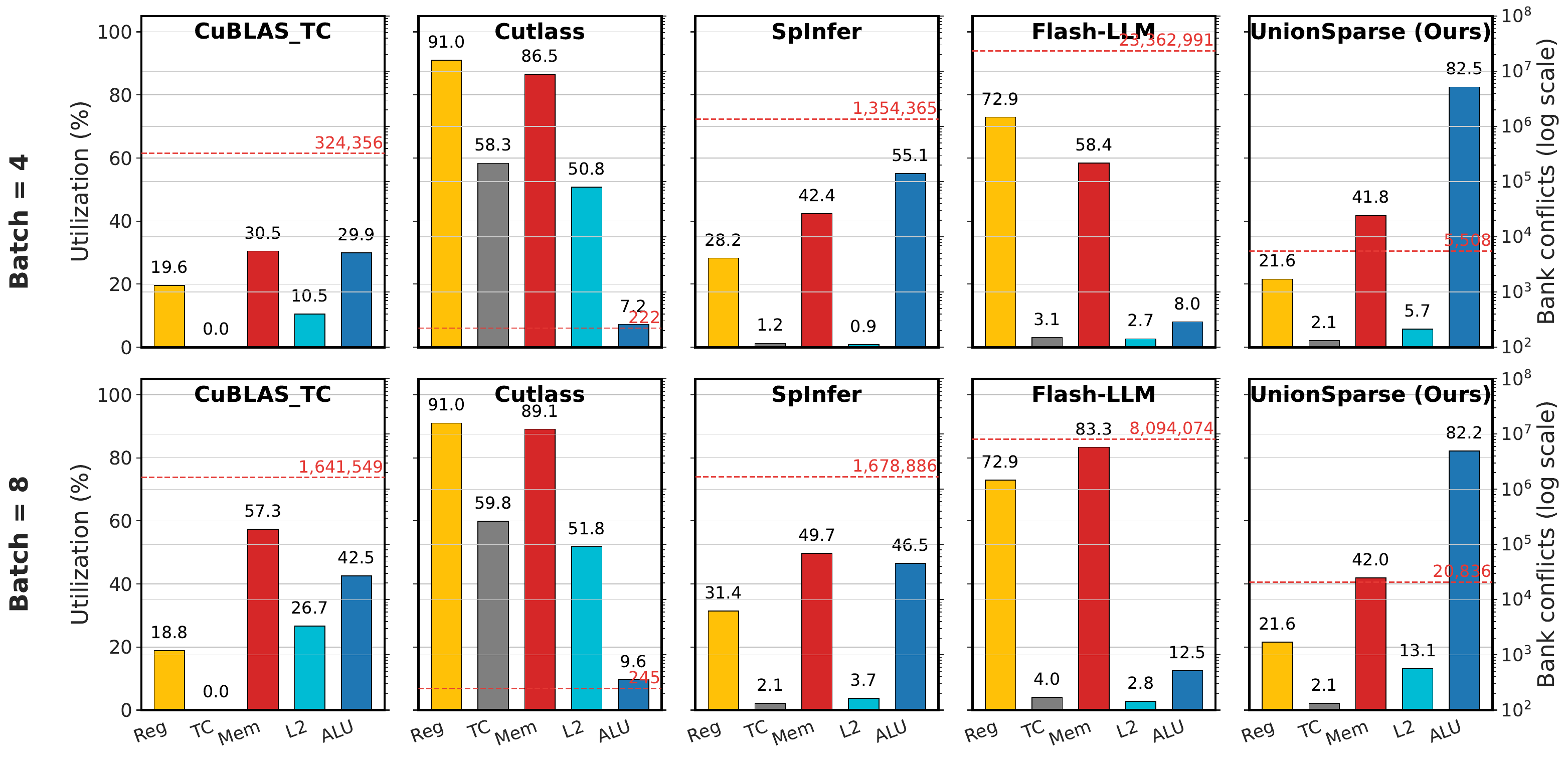}
  \caption{Kernel Utilization Breakdown under W4A4 Quantization for Two Representative MatMul Sizes (m=28672, k=7168, batch=8) and (m=12288, k=49152, batch=4).}
  \label{fig:int4-ncu}
\end{figure}

To better understand the source of the kernel-level gains, we further examine representative profiling metrics of different baselines under W4A4 quantization. Figure~\ref{fig:int4-ncu} reports the utilization breakdown for two representative matrix sizes, namely $(m=28672, k=7168, N=8)$ and $(m=12288, k=49152, N=4)$. We focus on register utilization, Tensor Core activity, memory-pipeline utilization, L2 utilization, ALU utilization, and shared-memory bank conflicts.

Several trends can be observed. First, among the sparse baselines, UnionSparse consistently achieves the highest ALU utilization, reaching 82.5\% at $N=4$ and 82.2\% at $N=8$, substantially higher than SpInfer and Flash-LLM. This indicates that UnionSparse is more effective at converting low-bit sparse computation into sustained on-chip execution, rather than stalling in metadata decoding or irregular sparse traversal. Second, UnionSparse also maintains higher L2 utilization than prior sparse baselines, indicating that its compressed representation, together with reorder, improves memory access regularity and enhances cache locality. Although dense baselines such as CUTLASS naturally exhibit stronger regularity on some metrics, they do not benefit from sparse arithmetic reduction and therefore serve mainly as dense-performance references.

A particularly important difference lies in bank conflicts. Compared with SpInfer and Flash-LLM, UnionSparse reduces shared-memory bank conflicts by orders of magnitude under both representative settings. This behavior is consistent with the design of IE-BME and LSPD: the shared union-mask representation reduces metadata indirection, while bitmap-guided four-stream gathering regularizes fragment reconstruction in shared memory and registers. In addition, the DW-PC pipeline decouples next-tile prefetch from current-tile processing, which helps avoid excessive concurrent pressure on the memory hierarchy. Together, these effects make the compressed sparse execution path substantially more hardware-friendly than prior sparse baselines.

These profiling results explain the performance trends observed in Fig.~\ref{fig:int4-speedup-by-n}. In the targeted low-$N$ regime, sparse low-bit execution is highly sensitive to metadata traffic, irregular accesses, and overlap inefficiency. By improving PMR on the representation side and reducing decode and staging overhead on the execution side, UnionSparse raises the effective compute intensity of sparse decode and allows the kernel to exploit the hardware more efficiently. This is precisely why its advantage is most pronounced in decode-oriented edge settings, where the cost of sparse metadata is hardest to amortize.

\begin{figure}[t]
\centering
\includegraphics[width=0.87\linewidth]{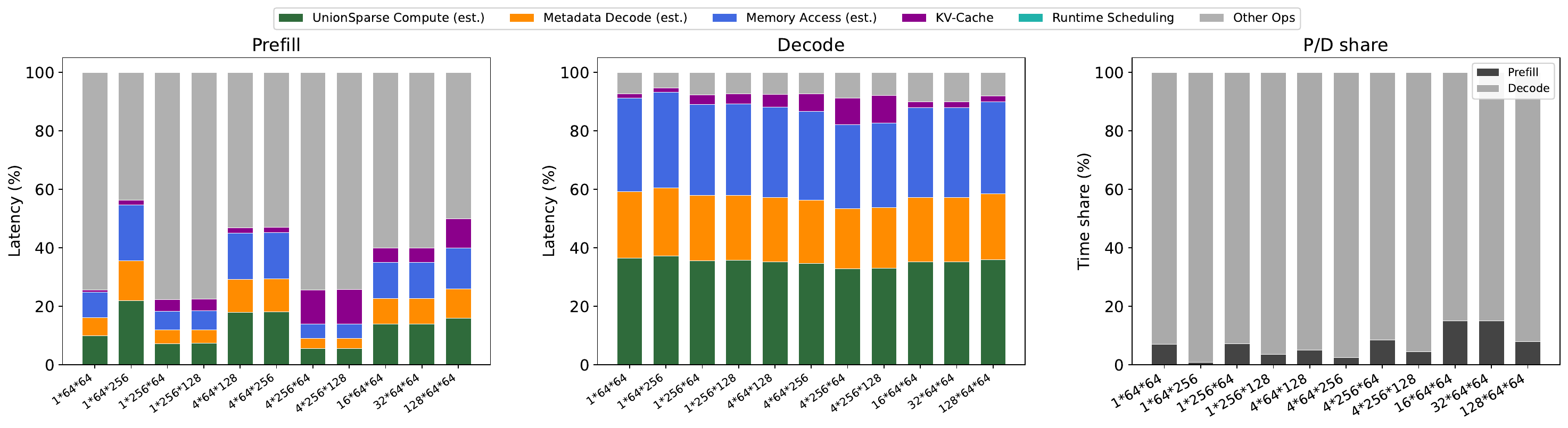}
\caption{End-to-end latency breakdown of UnionSparse during prefill and decode phases in the FasterTransformer pipeline.}
\label{fig:e2e_breakdown}
\end{figure}


\begin{figure*}[t]
  \centering
  \includegraphics[width=0.82\linewidth]{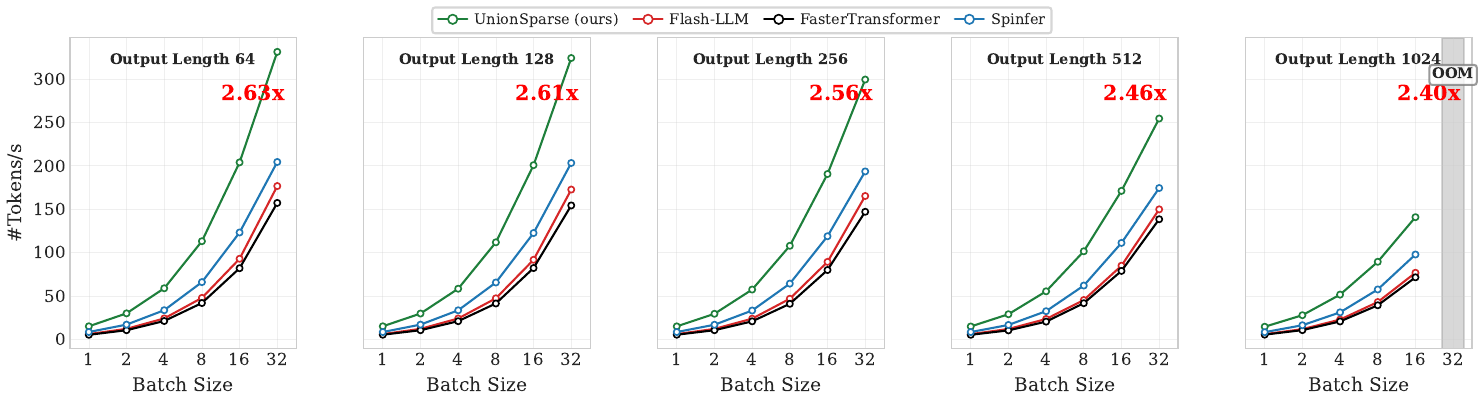}
  \caption{End-to-End Inference Performance of UnionSparse on OPT-13B with W4A4 Quantization.}
  \label{fig:e2e-w4a4}
\end{figure*}

\vspace{-10pt}
\subsection{End-to-End Model Evaluation}

We evaluate end-to-end inference performance of OPT-13B on Jetson AGX Orin under W4A4 quantization, as shown in Figure~\ref{fig:e2e-w4a4}. We first use Wanda~\cite{wanda} to prune the weight matrices to 50\% sparsity, and then apply Atom~\cite{zhao2024atom} to quantize the pruned weights to W4A4. After pruning and quantization, the resulting weights are converted from the dense layout to the UnionSparse format. We evaluate variable-length input sequences with an average input length of 10 tokens, which represents short-prompt edge inference scenarios.
Across all batch sizes and output lengths, \method{} consistently outperforms state-of-the-art frameworks, including Flash-LLM, FasterTransformer, and SpInfer.

In terms of throughput, \method{} achieves up to \textbf{{2.63$\times$}} speedup over FasterTransformer at output length 64.
As the output length increases, the relative speedup gradually decreases, reaching \textbf{{2.40$\times$}} at output length 1024.
This trend is consistent across different batch sizes, indicating that the benefit of \method{} is most pronounced in short-sequence and latency-sensitive decoding scenarios.

Compared to SpInfer, which already optimizes sparse execution, \method{} further improves performance by reducing payload extraction overhead and enhancing compute efficiency under low-bit sparse formats.
These gains are particularly evident in small-batch regimes, which are representative of real-world edge deployment settings.

    {
We profile the full FasterTransformer pipeline to decompose end-to-end latency into SpMM compute, metadata decoding, memory access, KV-cache, runtime scheduling, and non-SpMM operations.
In Figure~\ref{fig:e2e_breakdown}, each x-axis label follows $B=b,\ I=\ell_{\mathrm{in}},\ O=\ell_{\mathrm{out}}$, denoting batch size, input length, and output length, respectively. It shows that prefill is dominated by KV-cache writes and dense/non-SpMM operations, with SpMM contributing only $12\%$--$23\%$, so kernel-level speedup brings limited end-to-end gain.
In decode, however, $N=B$ is small and SpMM becomes the critical path, accounting for $69\%$--$75\%$ of per-token latency, while KV-cache and other operations are much smaller.
Thus, UnionSparse's system-level acceleration mainly comes from the decode phase, where reducing sparse weight access and SpMM latency directly translates to end-to-end speedup.
}

Despite substantial kernel-level improvements, the end-to-end speedup is lower than the peak kernel-level gains.
This is because SpMM kernel constitutes only part of the full inference pipeline.
Other components such as embedding lookup, attention scheduling, KV-cache management, and sampling remain unchanged and limit the achievable system-level speedup.

Moreover, as the output length increases, the proportion of non-SpMM operations grows, further diminishing the relative impact of SpMM acceleration.
This indicates a shift in the system bottleneck from compute-bound sparse kernels to pipeline-level coordination and memory-bound components.

Overall, these results demonstrate that \method{} effectively improves the efficiency of low-bit sparse decoding on edge GPUs, while also highlighting the importance of system-level co-design for achieving further end-to-end gains.

\begingroup

\subsection{End-to-End Energy Measurement}

\begin{table}[t]
\centering

\caption{End-to-end energy measurement of UnionSparse on OPT-13B under W4A4 quantization.}
\label{tab:e2e_energy}
\resizebox{\columnwidth}{!}{%

\begin{tabular}{cccc}
\toprule
Batch & Avg. Power (W) & Inc. Energy / Seq. (J) & Inc. Energy / Token (J) \\
\midrule
1 & 32.74 & 261.71 & 2.05 \\
2 & 32.96 & 135.14 & 1.06 \\
4 & 32.90 & 70.76  & 0.55 \\
8 & 32.82 & 38.69  & 0.30 \\
\bottomrule
\end{tabular}
}
\end{table}

Table~\ref{tab:e2e_energy} reports the end-to-end energy of UnionSparse on OPT-13B. The energy is measured with input length 64 and output length 128 using the on-board INA3221 power sensor through the hwmon/sysfs interface. The average power remains stable at around 33 W. As the batch size increases from 1 to 8, the incremental energy per generated token decreases from 2.05 J to 0.30 J, showing that batching amortizes weight access and runtime overheads and improves per-token energy efficiency.

\endgroup

\section{Conclusion}
This paper introduces UnionSparse, an index-efficient sparsity framework and SpMM kernel for low-bit sparse LLM inference on edge-GPU platforms.
By coupling Union Bitmap encoding with a decode-oriented SpMM kernel, UnionSparse reduces metadata overhead, regularizes sparse traversal, and improves the effective use of low-bit payloads.
The \vdir{} clarifies why these gains matter: in sparse decode, performance depends not only on how much data is removed, but also on how efficiently the remaining data can be indexed and consumed.
Across the measured INT4 benchmark and end-to-end framework, \method{} delivers clear kernel-level gains over dense and sparse baselines, with the strongest benefits appearing in low-$N$ and low-bit sparse decoding regimes where metadata costs are hardest to amortize.
More broadly, the results support co-designing sparse format layout and kernel mapping for latency and resource limited deployment.


\section*{Acknowledgment}
This work was supported in part by the Strategic Priority Research Program of the Chinese Academy of Sciences, Grant Nos. XDB0660101, XDB0660000, and XDB0660100, in part by the National Natural Science Foundation of China under Grant No. 62502489, in part by Jiangsu Provincial Natural Science Foundation under Grants BK20241818,BK20251815, and BK20250479,  in part by  Anhui Provincial Natural Science Foundation under Grant 2508085MF141, in part by the Studentsnnovation and Entrepreneurship Foundation of University of Scienceand Technology of China (USTC) under Grant CY2025X011B.

Generative AI tools were used during manuscript preparation only to assist with LaTeX formatting and editorial polishing. The authors reviewed and verified all AI-assisted edits and take full responsibility for the technical content, claims, experimental results, and conclusions of this article.

\bibliographystyle{IEEEtran}
\bibliography{unionsparse_refs}




\newcommand{\BioPhoto}[1]{%
  \includegraphics[width=0.72in,height=0.90in,clip,keepaspectratio]{#1}%
}
\makeatletter
\def\@IEEEBIOphotowidth{0.72in}
\def\@IEEEBIOphotodepth{0.90in}
\def\@IEEEBIOhangwidth{0.82in}
\def\@IEEEBIOhangdepth{0.90in}
\def\@IEEEBIOskipN{0pt}
\makeatother

\begin{IEEEbiography}[\BioPhoto{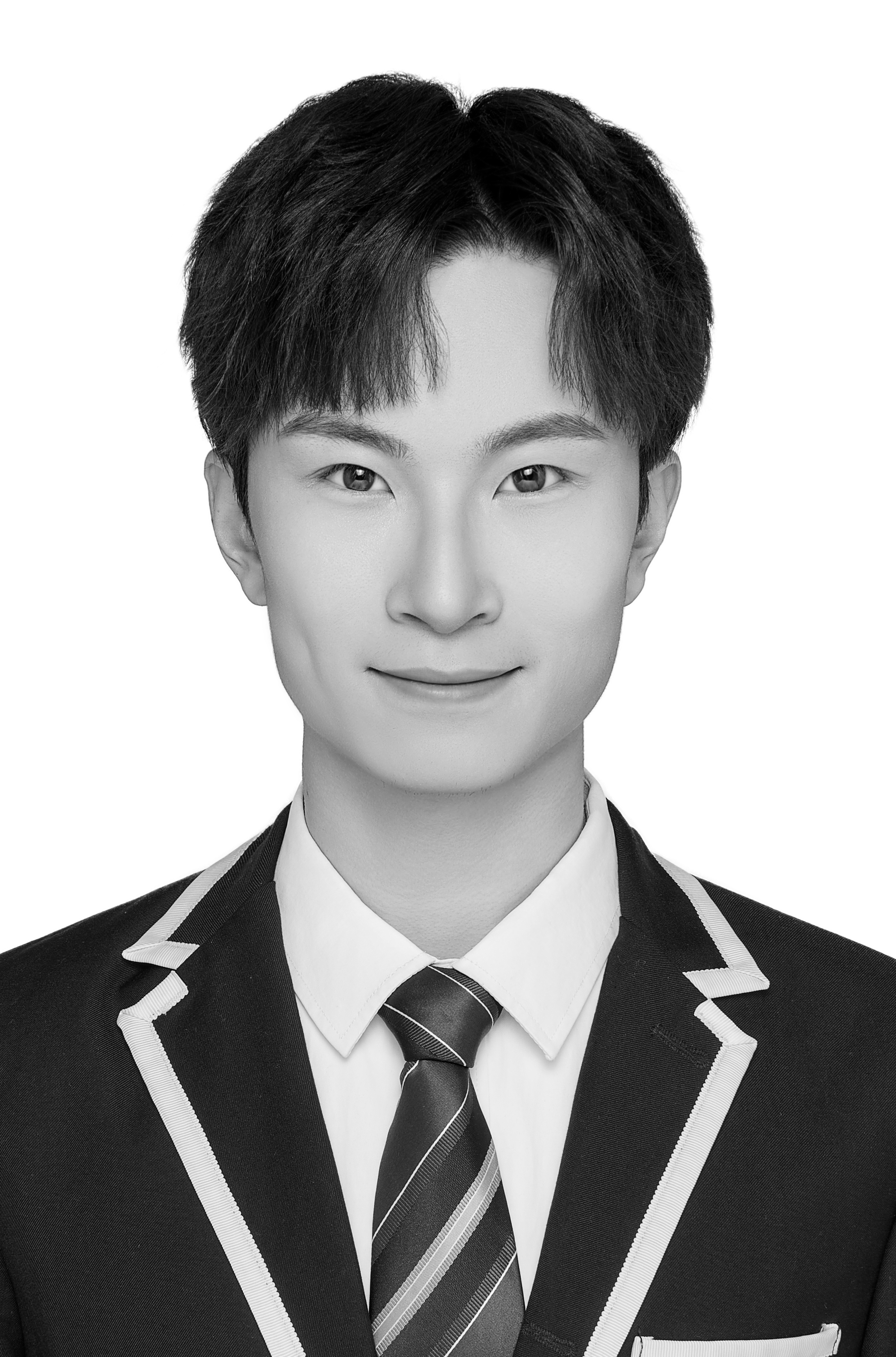}]{Tianhao Jiang}
 is currently pursuing the M.S. degree at the University of Science and Technology of China. He received the B.S. degree from East China Normal University. His research interests include efficient large language model inference in cloud and edge environments, distributed AI systems, FPGA-based acceleration, and performance optimization for heterogeneous computing platforms.
\end{IEEEbiography}

\begin{IEEEbiography}[\BioPhoto{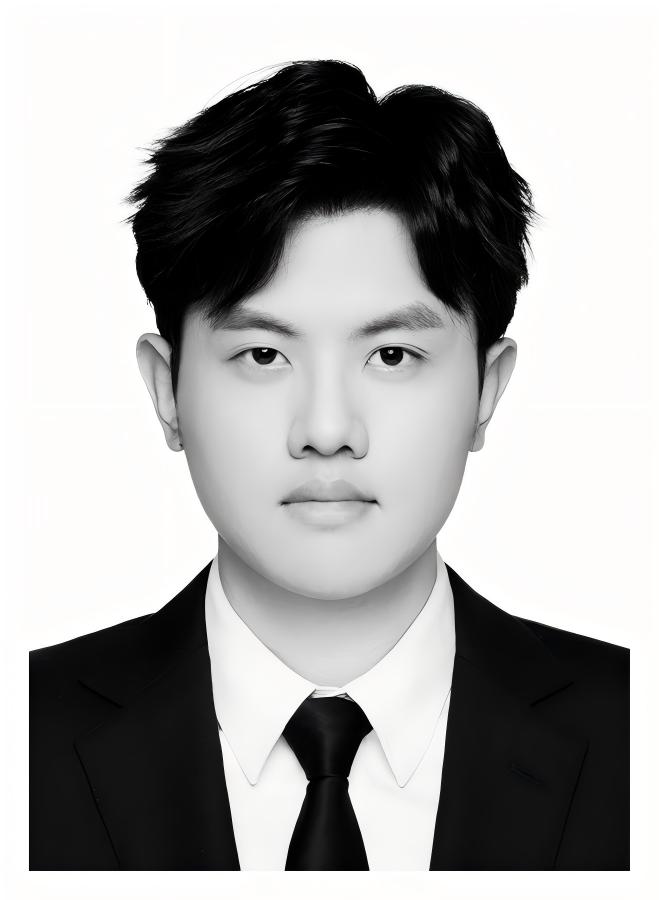}]{Hang Gu}
 received the B.E. degree in computer science and technology from Southeast University, Nanjing, China, in 2023. In the same year, he began his master's studies with the School of Computer Science and Technology, University of Science and Technology of China (USTC), Hefei, China. He is currently pursuing the Eng.D. degree with the School of Software Engineering, USTC. His research focuses on inference optimization and heterogeneous computing.
\end{IEEEbiography}

\begin{IEEEbiography}[\BioPhoto{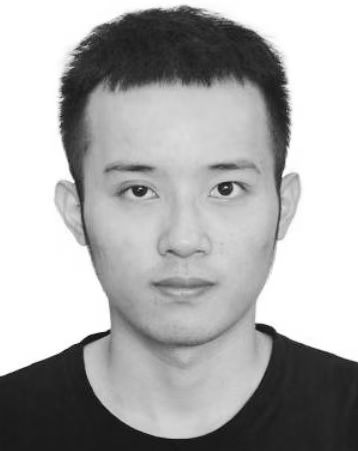}]{Teng Wang}
 is currently a research scientist with the Suzhou Institute for Advanced Research, University of Science and Technology of China. His research interests focus on algorithm-level and architecture-level acceleration for deep learning applications and AI-assisted architecture automation.
\end{IEEEbiography}

\begin{IEEEbiography}[\BioPhoto{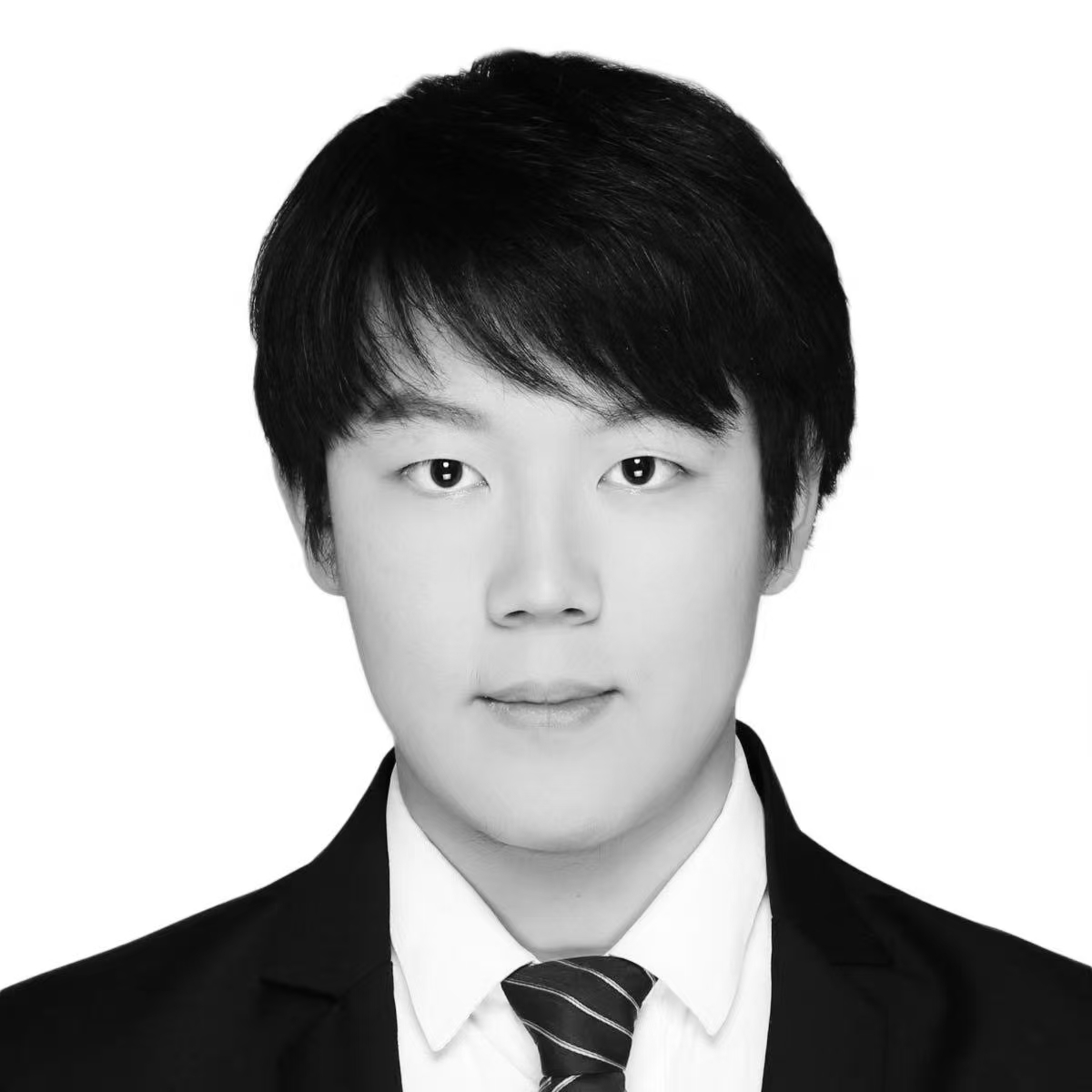}]{Qianyu Cheng}
 is currently a Ph.D. candidate in the School of Computer Science and Technology, University of Science and Technology of China. He received the bachelor's degree from the Yingcai Honors College, University of Electronic Science and Technology of China, in 2022. His research interests include CPU-FPGA heterogeneous systems, distributed accelerator systems, and their applications to large-scale analytical processing and multi-tenant cloud services.
\end{IEEEbiography}

\begin{IEEEbiography}[\BioPhoto{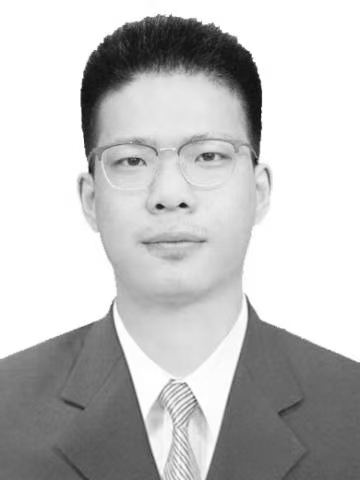}]{ZhenDong Zheng}
 received the B.S. degree from the School of Computer Science and Technology, University of Science and Technology of China, Hefei, China, in 2022, where he is currently pursuing the Ph.D. degree. His current research focuses on LLM accelerators and automated programming framework design.
\end{IEEEbiography}

\begin{IEEEbiography}[\BioPhoto{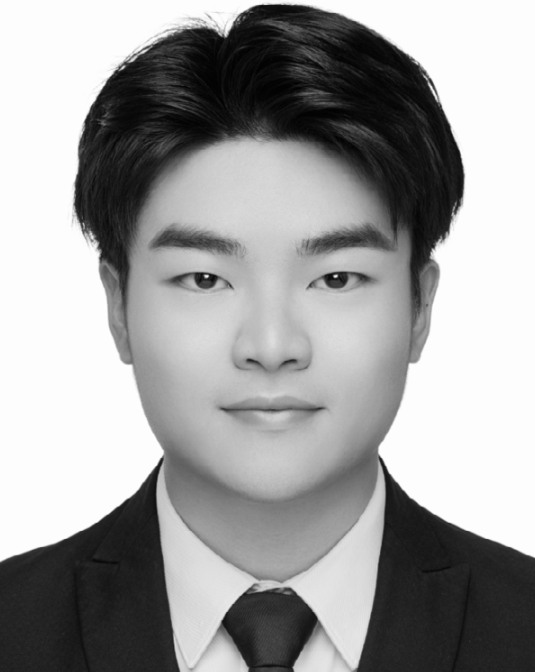}]{Cheng Tang}
 is currently pursuing the Ph.D. degree in software engineering with the University of Science and Technology of China, Hefei, China. His research interests include large language model inference, machine learning systems, and efficient AI computing. His research focuses on improving the efficiency, scalability, and deployability of large language models across heterogeneous and resource-constrained computing platforms.
\end{IEEEbiography}

\begin{IEEEbiography}[\BioPhoto{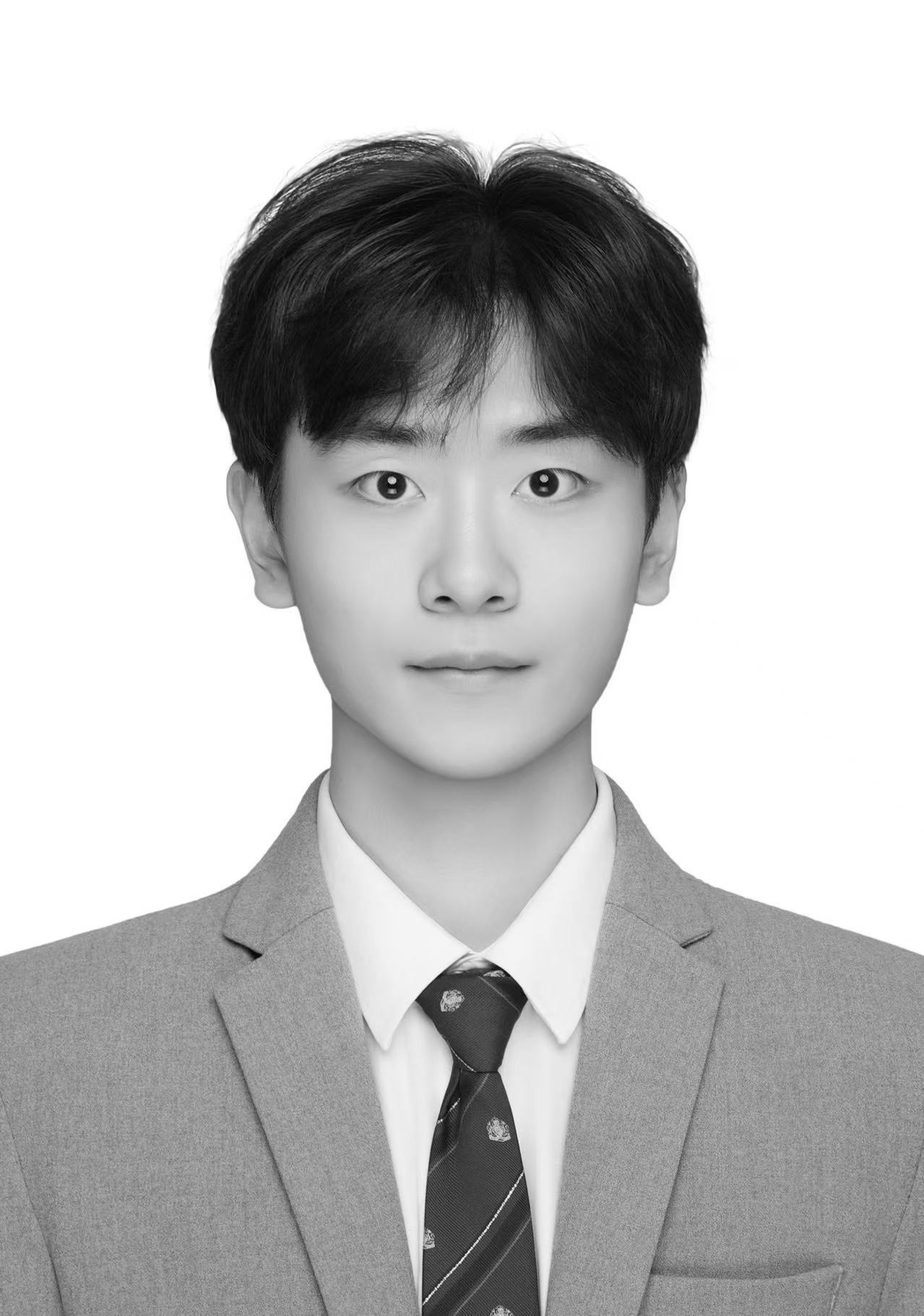}]{Qiyue Su}
 is currently pursuing the M.S. degree at the University of Science and Technology of China. He received the B.S. degree from Harbin Institute of Technology. His research interests include recommender systems, fully homomorphic encryption, and machine learning.
\end{IEEEbiography}

\begin{IEEEbiography}[\BioPhoto{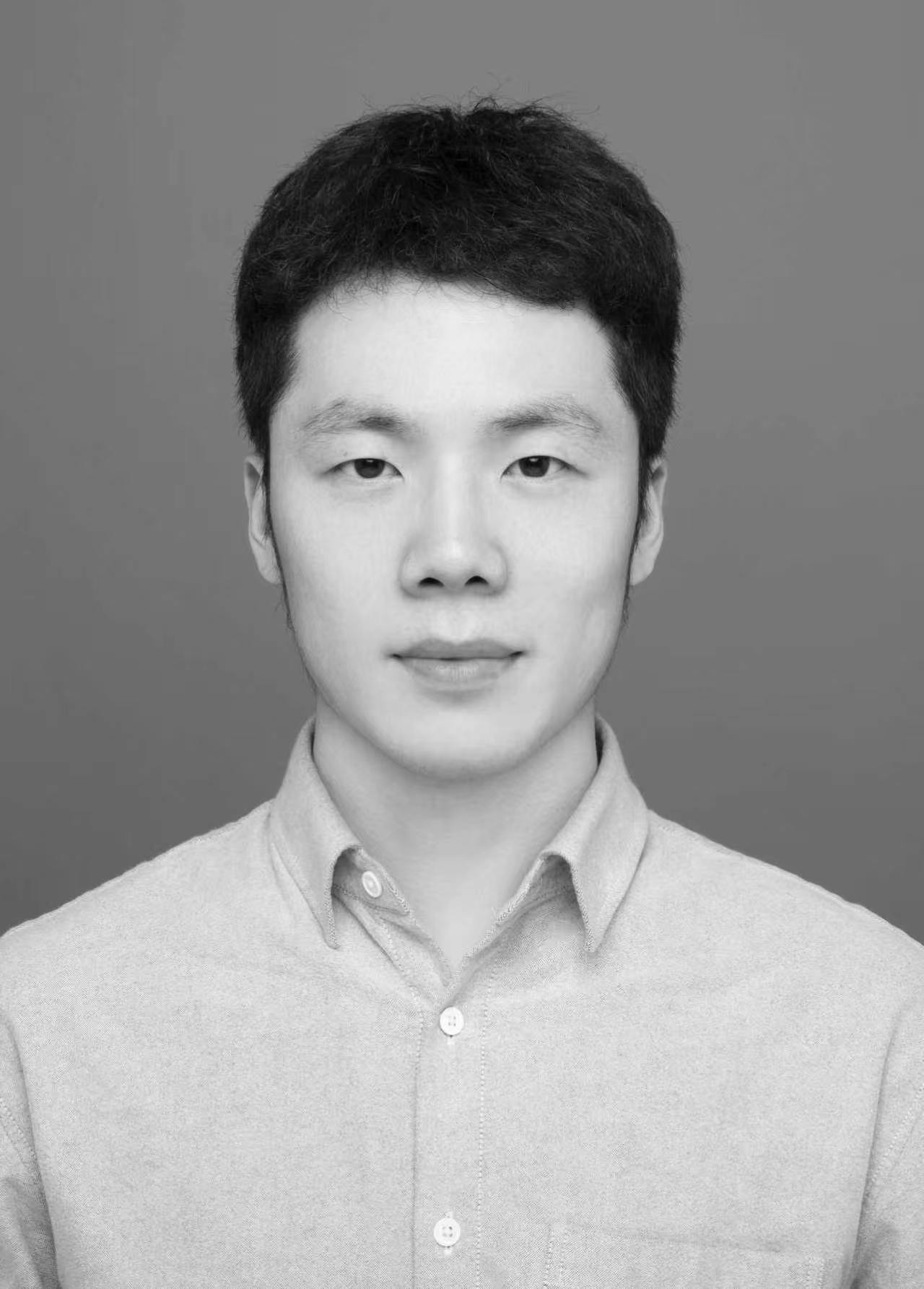}]{Wenqi Lou}
 (Member, IEEE) received the B.S. degree from Northwestern Polytechnical University, Xi'an, China, in 2018, and the Ph.D. degree in computer science from the University of Science and Technology of China, Hefei, China, in 2023. He is currently an Associate Researcher with the School of Software Engineering, University of Science and Technology of China. His current research interests include deep learning accelerators and FPGA-based acceleration.
\end{IEEEbiography}

\begin{IEEEbiography}[\BioPhoto{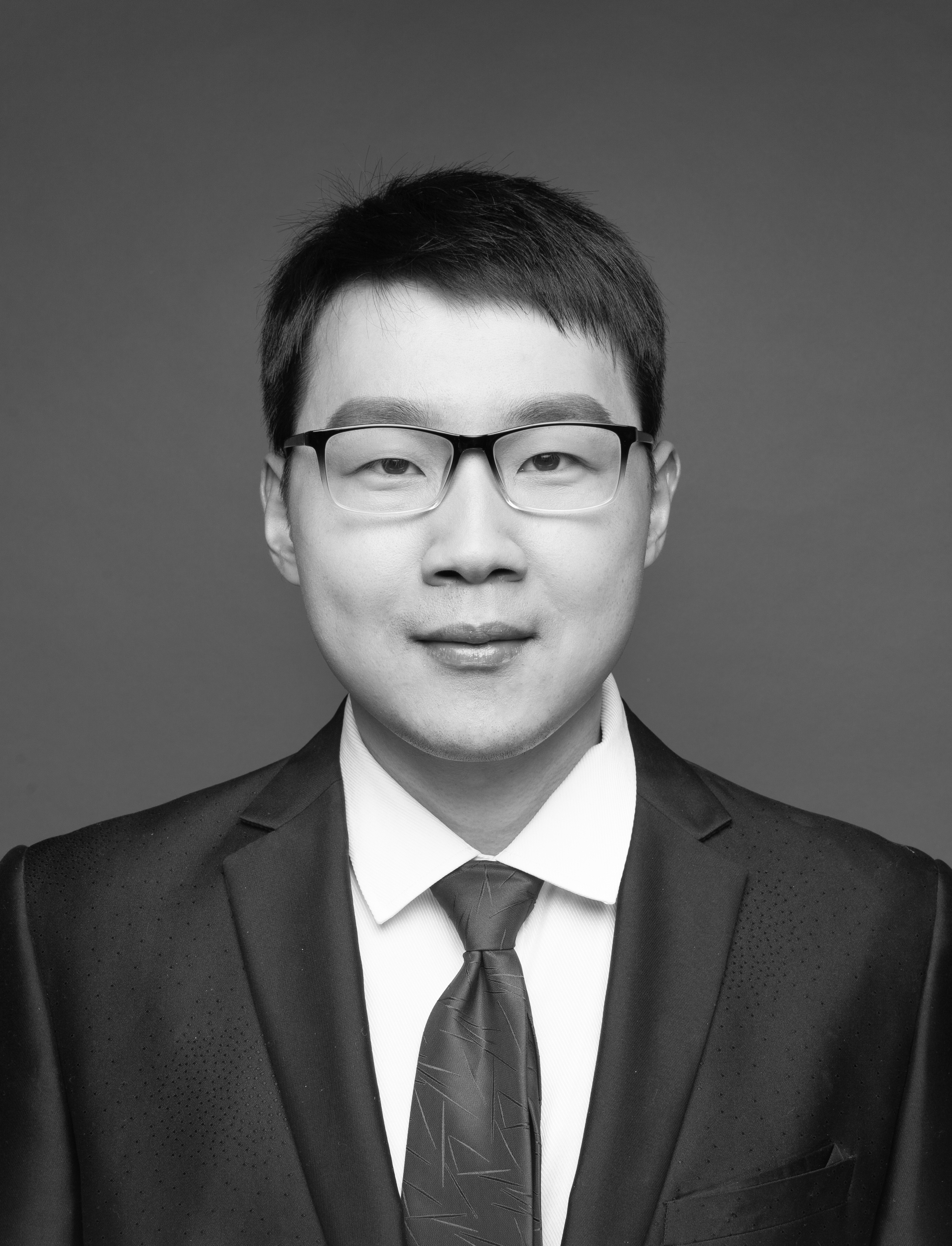}]{Lei Gong}
 received the Ph.D. degree in computer science from the University of Science and Technology of China in 2019. He is currently a research scientist with the School of Computer Science, University of Science and Technology of China. He examines various aspects of embedded systems, with a focus on performance, availability, flexibility, and energy efficiency. His research interests include machine learning accelerators and FPGA-based systems.
\end{IEEEbiography}

\begin{IEEEbiography}[\BioPhoto{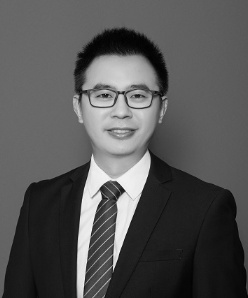}]{Chao Wang}
 is currently a Professor with the University of Science and Technology of China, Hefei, China. His research interests focus on multicore and reconfigurable computing. He serves as an Associate Editor of the ACM Transactions on Design Automation of Electronic Systems, the IEEE/ACM Transactions on Computational Biology and Bioinformatics, and Microprocessors and Microsystems. He is a Senior Member of the ACM, IEEE, and CCF.
\end{IEEEbiography}

\begin{IEEEbiography}[\BioPhoto{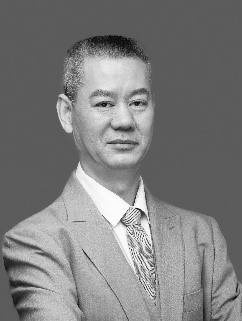}]{Xi Li}
 received the Ph.D. degree in computer science from the University of Science and Technology of China, Hefei, China, in 2003. He is currently a Professor with the University of Science and Technology of China, Hefei, China. He directs the research programs in the Embedded Systems Laboratory, examining various aspects of embedded systems, with a focus on performance, availability, flexibility, and energy efficiency. He has led several national key projects of China, several National High-Tech R\&D Program (863 Program) projects, and projects supported by the National Natural Science Foundation of China.
\end{IEEEbiography}

\begin{IEEEbiography}[\BioPhoto{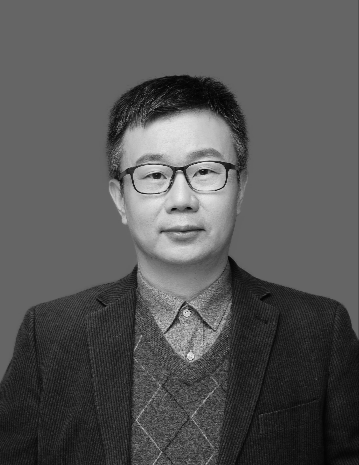}]{Xuehai Zhou}
 received the B.S., M.S., and Ph.D. degrees from the University of Science and Technology of China, Hefei, China, in 1987, 1990, and 1997, respectively. He is currently a Professor with the School of Computer Science, University of Science and Technology of China. His current research interests include various aspects of multicore and distributed systems.
\end{IEEEbiography}

\end{document}